\input harvmac
%\sequentialequations
%\draftmode
\overfullrule=0pt
\def\bigZ{Z\!\!\!Z}
\def\bigR{{\rm I}\!{\rm R}}
\def\muhyp{\mu_{\rm hyp}}
\def\Nequalstwo{{\rm v}}
\def\dmuone{d\mu^{(1)}}
\def\Cbar{\bar{\cal C}}
\def\Qonebar{\bar Q_1}
\def\Qtwobar{\bar Q_2}
\def\Aone{{A^{\scriptscriptstyle(1)}}}
\def\Atwo{{A^{\scriptscriptstyle(2)}}}
\def\dmuonephys{d\mu^{(1)}_{\rm phys}}
\def\K{{\cal K}}
\def\Kt{\tilde\K}
\def\thetabar{\bar\theta}
\def\Cf{{\cal C}_{\sst\cal N}}

\def\Pinfty{{\cal P}_{\sst\infty}}
\def\Atot{{\A_{\rm tot}}}

\def\uA{\,\lower 1.2ex\hbox{$\sim$}\mkern-13.5mu A}
\def\bigL{{\bf L}}

\def\zero{{\scriptscriptstyle(0)}}
\font\authorfont=cmcsc10 \ifx\answ\bigans\else scaled\magstep1\fi
%\divide\baselineskip by 7
%\multiply\baselineskip by 6
\divide\baselineskip by 10
\multiply\baselineskip by 9
%\advance\topskip by -1in\relax
\def\prenomat{\hbox{hep\hbox{-}th/9708036}}
\Title{$\prenomat$}{\vbox{\centerline{Supersymmetry and the
Multi-Instanton Measure
}}}
\centerline{\authorfont Nicholas Dorey}
\bigskip
\centerline{\sl Physics Department, University of Wales Swansea}
\centerline{\sl Swansea SA2$\,$8PP UK $\quad$ \tt n.dorey@swansea.ac.uk}
\bigskip
\centerline{\authorfont Valentin V. Khoze}
\bigskip
\centerline{\sl Department of Physics, Centre for Particle Theory, 
University of Durham}
\centerline{\sl Durham DH1$\,$3LE UK $\quad$ \tt valya.khoze@durham.ac.uk}
\bigskip
\centerline{and}
\bigskip
\centerline{\authorfont Michael P. Mattis}
\bigskip
\centerline{\sl Theoretical Division T-8, Los Alamos National Laboratory}
\centerline{\sl Los Alamos, NM 87545 USA$\quad$ \tt mattis@pion.lanl.gov}
\vskip .3in
\def\hf{{\textstyle{1\over2}}}
\def\H{{\cal H}}
\def\quarter{{\textstyle{1\over4}}}
\noindent
We propose explicit formulae for the integration measure on the moduli
space of \hbox{charge-$n$} ADHM multi-instantons in $N=1$ and $N=2$
supersymmetric gauge theories. The form of this measure is fixed
 by its (super)symmetries as well as the physical requirement of
clustering in the limit of large spacetime separation between
instantons. We test our proposals against known expressions for $n\le2$.
Knowledge of the measure for all $n$ allows us to revisit, and strengthen,
earlier $N=2$ results, chiefly: (1) For any number of flavors $N_F,$
we provide a closed formula for ${\cal F}_{n}$, the 
$n$-instanton contribution to the Seiberg-Witten prepotential,
 as a finite-dimensional collective coordinate
integral. This amounts to a solution, in
quadratures, of the Seiberg-Witten models, without appeal to
electric-magnetic duality.
%, thus 
%reducing the first-principles calculation of these numbers from the 
%Feynman path integral to quadrature.
 (2) In the conformal case $N_F=4,$
this means reducing to quadratures
 the previously unknown finite renormalization
that relates the microscopic and effective coupling constants,
$\tau_{\rm micro}$ and $\tau_{\rm eff}$.  
(3) Similar expressions are given for
the 4-derivative/8-fermion term in the gradient expansion of $N=2$
supersymmetric QCD.
%
%
%We continue our study of ADHM multi-instantons in $SU(2)$ gauge theory. 
%In the purely bosonic
%case, the collective coordinate integration measure is unknown for
%instanton numbers $n>2.$ Here we focus on  supersymmetric
%models, in which the multi-instanton
%collective coordinates have fermionic partners.
%For both $N=1$ and $N=2$ supersymmetry, we construct the measure for
%all $n$. As a corrollary, for any number of flavors,
%we give an explicit representation of
%${\cal F}_n$ (the $n$-instanton contribution to the Seiberg-Witten $N=2$
%prepotential) as a finite-dimensional collective coordinate integral.
%\looseness=-1
%\relax
\vskip .1in
\Date{\bf August 1997 } %replace this line
%by \draft  for preliminary versions or specify \draftmode at some point
\vfil\break

\lref\Seiberg{N. Seiberg, Phys. Lett. B206 (1988) 75. }
\lref\DineSeib{M. Dine and N. Seiberg, \it Comments
on higher derivative operators in some SUSY field theories\rm, hep-th/9705057.}
\lref\Henningson{M. Henningson, 
\it Extended Superspace, Higher Derivatives and 
$SL(2,\bigZ)$ Duality\rm, Nucl. Phys. B458 (1996) 445, hep-th/9507135.}
\lref\dWGR{B. de Wit, M.T. Grisaru and M. Rocek, \it Nonholomorphic
Corrections to the One-loop $N=2$ Super Yang-Mills Action\rm, 
Phys. Lett. B374 (1996) 297, hep-th/9601115.}
\lref\FFS{V.A. Fateev, I.V. Frolov and A.S. Shvarts,
 Nucl.~Phys.~\bf B154 \rm (1979) 1.}
\lref\dkmeight{N. Dorey, V.V. Khoze and M.P. Mattis, \it
Multi-instantons, three-dimensional gauge theory, and the
Gauss-Bonnet-Chern theorem\rm, hep-th/9704197.}
\lref\gilmore{See, e.g., R. Gilmore, \it Lie Groups, Lie Algebras and Some
of Their Applications\rm, Wiley-Interscience 1974.}
\lref\dadda{A. D'Adda and P. Di Vecchia, 
{\it Phys. Lett.} {\bf 73B} (1978) 162.}
\lref\oldyung{A. Yung, Nucl. Phys. B 344 (1990) 73.}
\lref\wessbagger{J. Wess and J. Bagger, {\it Supersymmetry and supergravity}, 
Princeton University Press, 1992.} 
\lref\Matone{M. 
Matone, {\it Instantons and recursion relations in $N=2$ SUSY gauge theory},
 Phys. Lett. B357 (1995) 342,    hep-th/9506102.  }
\lref\Osborntwo{H. Osborn,  Nucl. Phys. B140 (1978) 45.} 
\lref\GMO{P. Goddard, P. Mansfield, and H. Osborn,  Phys.~Lett.~\bf98B
\rm (1981) 59.}
\lref\NSVZ{ V. A. Novikov, M. A. Shifman, A. I. Vainshtein and
V. I. Zakharov, Nucl Phys. B229 (1983) 394; Nucl. Phys. B229 (1983)
407; Nucl. Phys. B260 (1985) 157. }
\lref\CWS{ N. H. Christ, E. J. Weinberg and N. K. Stanton, Phys
Rev D18 (1978) 2013. }
\lref\ADHM{M.  Atiyah, V.  Drinfeld, N.  Hitchin and
Yu.~Manin, Phys. Lett. A65 (1978) 185. }
\lref\Osborn{H. Osborn, Ann. Phys. 135 (1981) 373. }
\lref\CGTone{ E. Corrigan, P. Goddard and S. Templeton,
Nucl. Phys. B151 (1979) 93; \hfil\break
   E. Corrigan, D. Fairlie, P. Goddard and S. Templeton,
    Nucl. Phys. B140 (1978) 31.}
\lref\Fucito{F. Fucito and G. Travaglini, 
{\it Instanton calculus and nonperturbative relations
 in $N=2$ supersymmetric gauge theories}, 
 ROM2F-96-32, hep-th/9605215.}
\lref\dkmone{N. Dorey, V.V. Khoze and M.P. Mattis, \it Multi-instanton
calculus in $N=2$ supersymmetric gauge theory\rm, hep-th/9603136,
Phys.~Rev.~D54 (1996) 2921.}
\lref\dkmseven{N. Dorey, V.V. Khoze and M.P. Mattis, in preparation.}
\lref\dkmfive{N. Dorey, V.V. Khoze and M.P. Mattis, \it On $N=2$
supersymmetric QCD with $4$ flavors\rm,  hep-th/9611016, 
Nucl.~Phys.~\bf B492 \rm (1997) 607.}
\lref\dkmnine{N. Dorey, V.V. Khoze, M.P. Mattis,  M. Slater and W. Weir, 
\it Instantons, higher-derivative terms, and nonrenormalization theorems
in supersymmetric gauge theories\rm,
hep-th/9706007, Phys.~Lett.~\bf B \rm (in press).}
\lref\dkmtwo{N. Dorey, V.V. Khoze and M.P. Mattis, 
\it Multi-instanton check of the relation between the prepotential  
${\cal F}$ 
and the modulus $u$ in $N=2$ SUSY Yang-Mills theory\rm,
hep-th/9606199, Phys.~Lett.~B389 (in press, 26 December 1996).}
\lref\dkmthree{N. Dorey, V.V. Khoze and M.P. Mattis, 
\it A two-instanton test of the exact solution of $N=2$ supersymmetric  
QCD\rm,
hep-th/9607066, Phys.~Lett.~B388 (1996) 324.}
\lref\dkmfour{N. Dorey, V.V. Khoze and M.P. Mattis, \it Multi-instanton
calculus in $N=2$ supersymmetric gauge theory.
II. Coupling to matter\rm, hep-th/9607202, Phys.~Rev.~D54 (1996) 7832.}
\lref\HS{T. Harano and M. Sato, \it Multi-instanton
calculus versus exact results in $N=2$ supersymmetric QCD\rm,
hep-th/9608060}
\lref\Aoyama{H. Aoyama, T. Harano, M. Sato and S.Wada,
\it   Multi-instanton calculus in $N=2$ supersymmetric QCD\rm,
hep-th/9607076.}   
\lref\SWone{N. Seiberg and E. Witten, 
{\it Electric-magnetic duality, monopole
condensation, and confinement in $N=2$ supersymmetric Yang-Mills  
theory}, 
Nucl. Phys. B426 (1994) 19, (E) B430 (1994) 485  hep-th/9407087}
\lref\SWtwo{
N. Seiberg and E. Witten, 
{\it Monopoles, duality and chiral symmetry breaking
in $N=2$ supersymmetric QCD}, 
Nucl. Phys B431 (1994) 484 ,  hep-th/9408099}
\lref\BMSTY{T. Eguchi and S.K. Yang, 
{\it Prepotentials of $N=2$ supersymmetric gauge theories 
and soliton equations}, hep-th/9510183,
Mod.~Phys.~Lett.~\bf A11 \rm (1996) 131;
\hfil\break
G. Bonelli and M. Matone, 
    Phys. Rev. Lett. 76 (1996) 4107, hep-th/9602174; \hfil\break
 J. Sonnenschein, S. Theisen and S. Yankielowicz,
 Phys. Lett. 367B (1996) 145, hep-th/9510129;
\hfil\break
 N. Dorey, V. Khoze and M. Mattis, \it Multi-Instanton
Check of the Relation between the Prepotential $\cal F$ and the
Modulus $u$ in $N=2$ SUSY Yang-Mills Theory\rm,
hep-th/9606199,  
 Phys.~Lett.~\bf B390 \rm (1997) 205;
\hfil\break
F. Fucito and G. Travaglini, 
{\it Instanton calculus and nonperturbative relations
 in $N=2$ supersymmetric gauge theories}, 
 hep-th/9605215,
Phys.~Rev.~\bf D55 \rm (1997) 1099;\hfil\break
P.S. Howe and P. West, 
{\it Superconformal Ward identities and $N=2$ Yang-Mills theory},
hep-th/9607239,
Nucl.~Phys.~\bf B486 \rm (1997) 425.}
\lref\FPone{ D. Finnell and P. Pouliot,
{\it Instanton calculations versus exact results in $4$ dimensional 
SUSY gauge theories},
Nucl. Phys. B453 (95) 225, hep-th/9503115. }
\lref\tHooft{G. 't Hooft, Phys. Rev. D14 (1976) 3432; ibid.
D18 (1978) 2199.}
\def\frac#1#2{{ {#1}\over{#2}}}
\def\bose{{\rm bose}}

\def\fermi{{\rm fermi}}
\def\trtwo{\tr^{}_2\,}
\def\finv{f^{-1}}
\def\Ubar{\bar U}
\def\wbar{\bar w}

\def\abar{\bar a}
\def\bbar{\bar b}

\def\Deltabar{\bar\Delta}
\def\dalpha{{\dot\alpha}}
\def\dbeta{{\dot\beta}}
\def\dgamma{{\dot\gamma}}

\def\sst{\scriptscriptstyle}

\def\F{{\cal F}}

\def\A{{\cal A}}
\def\susy{supersymmetry}
\def\sigmabar{\bar\sigma}

\def\cl{{\,\rm cl}}
\def\lambdabar{\bar\lambda}

\def\psibar{\bar\psi}
\def\sqrtwo{\sqrt{2}\,}

\def\Qbar{\bar Q}
\def\susic{supersymmetric}

\def\vhiggs{{\rm v}}

\def\vhiggsa{{{\cal A}_{\sst00}}}
\def\vbarhiggs{\bar{\rm v}}

\def\C{{\cal C}}

\def\new{{\scriptscriptstyle\rm new}}

\def\uX{\,\lower 1.2ex\hbox{$\sim$}\mkern-13.5mu X}
\def\uQ{\,\lower 1.2ex\hbox{$\sim$}\mkern-13.5mu Q}
\def\uQtilde{\,\lower 1.2ex\hbox{$\sim$}\mkern-13.5mu \tilde Q}
\def\uD{\,\lower 1.2ex\hbox{$\sim$}\mkern-13.5mu {\rm D}}

\def\uF{\,\lower 1.2ex\hbox{$\sim$}\mkern-13.5mu F}
\def\uW{\,\lower 1.2ex\hbox{$\sim$}\mkern-13.5mu W}
\def\uWbar{\,\lower 1.2ex\hbox{$\sim$}\mkern-13.5mu {\overline W}}
\def\uPhibar{\,\lower 1.2ex\hbox{$\sim$}\mkern-13.5mu {\overline \Phi}}

\def\uV{\,\lower 1.2ex\hbox{$\sim$}\mkern-13.5mu V}
\def\uv{\,\lower 1.0ex\hbox{$\scriptstyle\sim$}\mkern-11.0mu v}
\def\uPsi{\,\lower 1.2ex\hbox{$\sim$}\mkern-13.5mu \Psi}
\def\uPhi{\,\lower 1.2ex\hbox{$\sim$}\mkern-13.5mu \Phi}
\def\uchi{\,\lower 1.5ex\hbox{$\sim$}\mkern-13.5mu \chi}
\def\uchitilde{\,\lower 1.5ex\hbox{$\sim$}\mkern-13.5mu \tilde\chi}
\def\Psibar{\bar\Psi}
\def\uPsibar{\,\lower 1.2ex\hbox{$\sim$}\mkern-13.5mu \Psibar}
\def\upsi{\,\lower 1.5ex\hbox{$\sim$}\mkern-13.5mu \psi}
\def\uq{\,\lower 1.5ex\hbox{$\sim$}\mkern-13.5mu q}
\def\uqtilde{\,\lower 1.5ex\hbox{$\sim$}\mkern-13.5mu \tilde q}
\def\psibar{\bar\psi}
\def\upsibar{\,\lower 1.5ex\hbox{$\sim$}\mkern-13.5mu \psibar}
\def\upsibarzero{\,\lower 1.5ex\hbox{$\sim$}\mkern-13.5mu \psibar^\zero}
\def\ulambda{\,\lower 1.2ex\hbox{$\sim$}\mkern-13.5mu \lambda}
\def\ulambdabar{\,\lower 1.2ex\hbox{$\sim$}\mkern-13.5mu \lambdabar}
\def\ulambdabarzero{\,\lower 1.2ex\hbox{$\sim$}\mkern-13.5mu \lambdabar^\zero}
\def\ulambdabarnew{\,\lower 1.2ex\hbox{$\sim$}\mkern-13.5mu \lambdabar^\new}
\def\D{{\cal D}}
\def\M{{\cal M}}
\def\N{{\cal N}}
\def\Dslash{\,\,{\raise.15ex\hbox{/}\mkern-12mu \D}}
\def\Dbarslash{\,\,{\raise.15ex\hbox{/}\mkern-12mu {\bar\D}}}
\def\delslash{\,\,{\raise.15ex\hbox{/}\mkern-9mu \partial}}
\def\delbarslash{\,\,{\raise.15ex\hbox{/}\mkern-9mu {\bar\partial}}}

\def\hf{{\textstyle{1\over2}}}
\def\quarter{{\textstyle{1\over4}}}
\def\eighth{{\textstyle{1\over8}}}

\def\xibar{\bar\xi}

\def\uAcl{\,\lower 1.2ex\hbox{$\sim$}\mkern-13.5mu A^{}_{\cl}}
\def\uAbarcl{\,\lower 1.2ex\hbox{$\sim$}\mkern-13.5mu A_{\cl}^\dagger}

\newsec{Introduction}
Since their discovery, instantons have continued to play a central
role in our understanding of non-perturbative effects in gauge
theory. They are unique in
providing  non-perturbative effects which can nevertheless be
calculated systematically in the semiclassical limit. Usually the main
technical obstacle
in such a calculation is to determine the correct quantum measure for
integration over the collective coordinates of the instanton. In the
case of a single instanton (topological charge $n=1$), 
this measure was determined by \hbox{'t Hooft} in the classic paper
\tHooft. In general, the complete measure is unknown
for $n>1$ although, as we will discuss below, an important part of the
answer has been determined for the case $n=2$. In this 
paper we will propose explicit expressions for the $n$-instanton
measure for the gauge group $SU(2)$, 
both in theories with $N=1$ supersymmetry (SUSY) 
(see Eqs.~(2.23) and (2.54) below)  and also in theories with 
$N=2$ SUSY (see Eqs.~(3.19) and (3.27) below). 
As non-trivial tests of our proposals, we will show that they
reproduce known results in the one- and two-instanton sectors.  

Instanton effects are especially prominent in the $N=2$ theories, 
where they provide the only corrections to the
holomorphic prepotential $\F$ beyond one-loop in perturbation 
theory \refs{\Seiberg\SWone-\SWtwo}.
Our expression given below for the $N=2$ multi-instanton measure,
combined with earlier results from Ref.~\dkmfour, yields 
a closed expression for the $n$-instanton
contribution to $\F$ as a finite-dimensional integral
over bosonic and fermionic collective coordinates.  Related expressions
will be given for the 4-derivative/8-fermion term in the
gradient expansion of $N=2$ \susic\ QCD.

The relevant field configurations in the $N=1$ and $N=2$ theories
discussed above are supersymmetric extensions of the general 
multi-instanton solutions constructed by Atiyah, Drinfeld, Hitchin and
Manin (ADHM) \ADHM. The ADHM construction reduces the problem of
solving the self-dual Yang-Mills equation to that of solving purely
algebraic equations. Despite this major simplification, the resulting
algebraic constraints are highly non-linear and cannot be solved
explicitly for general $n$. In calculating physical quantities 
the basic aim is to integrate over 
the solution space of the ADHM constraints (modulo an internal $O(n)$
redundancy described below).  
It is often assumed that knowledge of an explicit
solution of the constraints
is a prerequisite for constructing the integration measure;
 this is the source of the widely-held belief that progress is
impossible beyond $n=2$.\foot{In fact a solution of the ADHM
constraints for $n=3$ was presented in \CWS, but the complexity
of the resulting expressions is such that no subsequent progress has
been made for this case.}

 In this paper we will adopt  
an alternative approach to this problem: we will define the collective
coordinate integral as an 
integral over a larger set of variables with the supersymmetrized
ADHM constraints
imposed by  $\delta$-functions under the integral sign. The problem then
is shifted to that of determining the correct algebraic form which
sits inside the integrand (including the Jacobians induced by the
\hbox{$\delta$-functions}). 
 The key simplification provided by SUSY which
makes this determination possible stems from the fact that 
the $N=1$ and $N=2$ algebras can be   realized as
transformations of the instanton moduli \refs{\NSVZ,\dkmfour}; in its most
symmetric form,  the integration measure must then
be invariant under these transformations. (This simplification is
in addition to the well-known observation that in a \susic\ theory,
the small-fluctuations 't Hooft determinants cancel in a self-dual
background between bosonic and fermionic excitations \dadda.)
This
invariance requirement, together with the other symmetries of the
problem, turns out to be sufficient to fix the
multi-instanton measure up to a
multiplicative constant for each $n$. 
These constants are then determined by induction, up to a single overall
normalization, by the
clustering property of the measure in the dilute gas limit where the
separation between instantons is much larger than their scale sizes.   
Finally, this single remaining normalization is fixed by comparison to 
't Hooft's result in the
one-instanton sector. As mentioned above, comparison to known expressions
in the two-instanton sector \refs{\Osborn,\dkmone}
then yields an independent test of our proposals.
It would be interesting to see if the collective coordinate measures
given below
can be deduced in an alternative way familiar from the study
of monopoles, by constructing the (supersymmetrized) volume form
from the hyper-K$\ddot{\rm a}$hler metric 2-form on the ADHM space.

The paper is organized as follows. Section 2 is devoted to the case of
$N=1$ \susic\ $SU(2)$ gauge theory. After a brief review of our ADHM
and SUSY conventions, we write down our Ansatz for the collective
coordinate integration measure, which is invariant not only under SUSY,
but also under the internal $O(n)$ symmetry mentioned above. Much of
 this section is devoted to  analyzing  the clustering
requirement, which fixes the overall normalization constants of the
$n$-instanton measure by induction in $n$. Clustering together with
zero-mode counting turns out to be
a highly restrictive requirement, ruling out virtually all other conceivable
SUSY- and $O(n)$-invariant forms for the measure. At the end of Sec.~2 we
show that for $n=2$ 
our postulated measure is equivalent to the known first-principles
2-instanton measure constructed in \refs{\Osborn,\dkmone}, in which
neither the SUSY nor the  $O(2)$ invariance is  manifest
(instead these symmetries are ``gauge-fixed''). In Sec.~3 we
repeat all these steps for the case of $N=2$ \susic\ $SU(2)$ gauge theory.
\hbox{Section 4}  briefly discusses
 the incorporation of matter supermultiplets, which poses no problems. 

In Sec.~5 we revisit the Seiberg-Witten prepotential $\F,$ and give
the  explicit collective coordinate integral formula for $\F_n$, the
$n$-instanton contribution to $\F$, valid for any number of flavors $N_F$.
This is tantamount to an all-instanton-orders  solution in quadratures of the
Seiberg-Witten models, without appeal to electric-magnetic duality. (For
 analogous multi-instanton
solutions of certain 2-dimensional and 3-dimensional models, see
for instance
Refs.~\FFS\ and \dkmeight, respectively.)
While
such a solution may be of purely academic interest for $N_F\le3$ in light
of the exact results of \refs{\SWone,\SWtwo}, for the
conformal case $N_F=4$ it gives something new: the all-instanton-orders
relation between the microscopic and effective coupling constants,
$\tau_{\rm micro}$ and $\tau_{\rm eff}$, which are known to be
inequivalent \refs{\dkmfour,\dkmfive}.  So far as we currently understand,
$SL(2,\bigZ)$ duality by itself cannot give this relation, as it operates
only at the level of $\tau_{\rm eff}$. Using the results of \dkmnine,
we give similar expressions for the pure-holomorphic and 
pure-antiholomorphic contributions to the 4-derivative/8-fermion term
in the gradient expansion along the Coulomb branch of $N=2$
\susic\ QCD. Section 5 concludes with comments about how one actually
performs these collective coordinate
integrations, although explicit calculational progress
along these lines is deferred to future work.

\newsec{The $N=1$ \susic\ collective coordinate integration measure}
\subsec{ADHM and SUSY review}

The basic object in the ADHM construction \ADHM\ of self-dual $SU(2)$  
gauge
fields of topological number $n$ is an $(n+1)\times n$  
quaternion-valued
matrix $\Delta_{\lambda l}(x)$, which is a linear function of the  
space-time
variable $x\,$:\foot{We use quaternionic notation $x=x_{\alpha\dalpha}
=x_n\sigma^n_{\alpha\dalpha},$ $\abar=\abar^{\dalpha\alpha}=\abar^n
\sigmabar_n^{\dalpha\alpha}$, $b=b_\alpha^{\ \beta},$ etc., where 
$\sigma^n$ and $\sigmabar^n$ are the spin matrices of Wess and Bagger
\wessbagger. 
See  Ref.~\dkmone\
for a self-contained introduction to the ADHM construction including a
full account of our ADHM and SUSY conventions.
We also set the coupling constant $g=1$ throughout, except, for clarity, in
the Yang-Mills instanton action $8\pi^2n/g^2.$
}
\eqn\Deltapostulate{\Delta_{\lambda l}
\ =\ a_{\lambda l}\ +\
b_{\lambda l}\,x\ ,\quad
0\le\lambda\le n\ ,\ \ 1\le l\le n\ .}
The gauge field $v_m(x)$ is then given by (displaying color indices)
\eqn\adhmansatz{v_m{}^\dalpha{}_\dbeta\ =\ 
\Ubar_\lambda^{\dalpha\alpha}\partial_m
U_{\lambda\,\alpha\dbeta}\ ,}
where the quaternion-valued vector $U_\lambda$ lives in the $\perp$  
space
of $\Delta\,$:
\eqna\perpspace
$$\eqalignno{\Deltabar_{l\lambda}\,U_{\lambda}
\ &=\ \Ubar_\lambda\,\Delta_{\lambda l}\ =\ 0\ ,&\perpspace a
\cr
\Ubar_\lambda U_\lambda\ &=\ 1\ .&\perpspace b}$$
It is easy to show that self-duality of the field strength $v_{mn}$ is
equivalent to the quaternionic condition
\eqn\quatcond{\Deltabar^{\dbeta\beta}_{k\lambda}\,\Delta^{}_{\lambda l\,
\beta\dalpha}\ =\ (\finv)_{kl}\,\delta^\dbeta{}_\dalpha}
for some scalar-valued $n\times n$ matrix $f$;  Taylor  expanding
in $x$ then gives
\eqna\crucial
$$\eqalignno{\abar a\ &=\ (\abar a)^T\ \propto\  
\delta^\dbeta{}_\dalpha\
,&\crucial a
\cr
\bbar a\ &=\ (\bbar a)^T\ ,&\crucial b\cr
\bbar b\ &=\ (\bbar b)^T\  \propto\ \delta_\alpha{}^\beta\
.&\crucial c}$$
Here the $\scriptstyle T$ stands for transpose in the ADHM indices
$(\lambda, l, \hbox{etc.})$ only, whereas an overbar indicates conjugation
in both the ADHM and the quaternionic spaces.

Following Refs.~\refs{\CGTone,\CWS,\dkmone}, we will work in a representation
in which $b$ assumes a simple canonical form, namely
\eqn\bcanonical{
a_{\alpha\dalpha}\ =\  
\pmatrix{w_{1\alpha\dalpha}&\cdots&w_{n\alpha\dalpha}
\cr{}&{}&{}\cr
{}&a'_{\alpha\dalpha}&{}\cr{}&{}&{}}\quad,\qquad
b_\alpha^{\ \beta}\ =\ 
\pmatrix{0&\cdots&0 \cr \delta_\alpha{}^\beta & \cdots & 0 \cr
\vdots & \ddots & \vdots \cr 0 & \cdots &
\delta_\alpha{}^\beta}}
Thanks to this simple form for $b$, 
the constraint \crucial c is now automatically satisfied, while
 \crucial b  reduces to the symmetry
condition on the $n\times n$ submatrix $a'\,$:
\eqn\symcond{a'\ =\ a^{\prime T}}
Note that there is an  $O(n)$ group of transformations
on $\Delta(x)$
which preserves this canonical form for $b$ as well as the ADHM conditions
\crucial{}, but acts nontrivially on $a\,$:
\def\dmunphys{d\mu^{(n)}_{\rm phys}}
\def\dmutwophys{d\mu^{(2)}_{\rm phys}}
\def\dmuonephys{d\mu^{(1)}_{\rm phys}}
\def\dmun{d\mu^{(n)}}

\def\dmunone{d\mu^{(n-1)}}
\def\VolOn{{\rm Vol}\big(O(n)\big)}

\eqn\trans{\Delta\,\rightarrow\
\pmatrix{1&0&\cdots&0\cr0&{}&{}&{}\cr\vdots&{}&R^T&{}\cr0&{}&{}&{}\cr}
\cdot\Delta\cdot R\ ,\quad f\rightarrow R^T\cdot f\cdot R\ ,
\quad U_0\rightarrow U_0\ ,\quad U_l\rightarrow R_{kl}U_k}
Here $R$ is an $O(n)$ matrix whose elements are independent of $x$ and
act by scalar multiplication on the quaternions.
{}From \adhmansatz\ one sees that these transformations do not affect the
gauge field $v_m$. Hence, the physical moduli space, ${\rm M}_{\rm
phys}$, of gauge-inequivalent self-dual gauge configurations is the
quotient of the space ${\rm M}$ of all 
solutions of the constraints (2.5) which have
the canonical form (2.6), by the symmetry group $O(n)$: 
\eqn\modspace{{\rm M}_{\rm phys} \  =\ {{\rm M} \over O(n)}}

Let us count the number of physical degrees of freedom represented by
the collective coordinate matrix $a$. Since the elements of $a$ are
quaternions, it contains $4(n+\hf n(n+1))$ real
scalar degrees of freedom thanks
to \symcond. The ADHM constraint \crucial a then imposes ${3\over2}n(n-1)$
conditions on the upper-triangular traceless quaternionic elements of $\abar
a,$ while the modding out by the 
$O(n)$ action \trans\ subtracts an additional $\hf n(n-1)$ degrees of freedom.
This leaves
\eqn\arith{4(n+\hf n(n+1))\ -\ \textstyle{{3\over2}n(n-1)}\ -\
\hf n(n-1)\ =\ 8n}
physical degrees of freedom for the multi-instanton with topological number
$n$. This is the correct number: 
in the limit of $n$ widely separated (i.e., distinguishable)
instantons it properly accounts for $4n$ positions, $3n$
iso-orientations, and $n$ instanton scale sizes. 

In an $N=1$ \susic\ theory the gauge field $v$ is accompanied by
a gaugino $\lambda$. By the index theorem, the
zero modes of $\lambda$
should comprise $4n$ Grassmann (i.e., anticommuting)
degrees of freedom. These adjoint fermion zero modes have the form
\CGTone
\eqn\lambdazm{(\lambda_\alpha)^\dbeta{}_\dgamma\ =\ 
\Ubar^{\dbeta\gamma}\M_\gamma f\,\bbar\, U_{\alpha\dgamma}\ -\
\Ubar^\dbeta{}_\alpha \,bf\M^{\gamma T}U_{\gamma\dgamma}\ .}
We suppress ADHM matrix indices ($\lambda,$ $l$, etc.)
but exhibit color (dotted) and Weyl
(undotted) indices for clarity. Here $\M$ is an $x$-independent
Weyl-spinor-valued $(n+1)\times n$ matrix:
\eqn\Mdef{\M^\gamma\ =\ \pmatrix{\mu_1^\gamma&\cdots&\mu_n^\gamma
\cr{}&{}&{}\cr
{}&\M^{\prime\gamma}&{}\cr{}&{}&{}}}
The 2-component Dirac equation in the background of the ADHM
multi-instanton \adhmansatz\ is equivalent to
 the following linear constraints on $\M_\gamma$  \CGTone:
\eqna\zmcons
$$\eqalignno{\abar^{\dalpha\gamma}\M_\gamma\ &=\ -\M^{\gamma T}a_\gamma
{}^\dalpha\ ,&\zmcons a
\cr
\M_\gamma^{\prime T}\ &=\ \M'_\gamma\ .&\zmcons b}$$
This leaves
\eqn\fermcount{2\big(n+\hf n(n+1)\big)\ -\ n(n-1)\ =\ 4n}
degrees of freedom in $\M,$ as is needed. Under $O(n)$, $\M$ transforms
just like $\Delta(x)$, Eq.~\trans.

Next we review the \susic\ properties of the collective coordinate
matrices $a$ and $\M$ \dkmfour. 
 As the relevant
field configurations $v_m$ and $\lambda_\alpha$
obey equations of motion which are manifestly
supersymmetric, any non-vanishing action of 
the  supersymmetry generators on a 
particular solution necessarily yields another solution. It follows
that the ``active''
supersymmetry transformations of the fields must be equivalent (up  
to a
gauge transformation) to certain ``passive'' transformations of 
the $8n$ independent bosonic and $4n$ independent
fermionic collective coordinates which parametrize
the superinstanton solution. As originally noted in \NSVZ, 
physically relevant quantities such as the
saddle-point action of the superinstanton must be constructed out of
\susic\ invariant combinations of the collective coordinates.   

An especially attractive feature of the ADHM
construction is that the 
 supersymmetry algebra can actually be realized  
directly as transformations of 
the highly over-complete (order $n^2$, rather than order $n$) set of collective
coordinates $a$ and $\M$.
Under
an infinitesimal \susy\ transformation $\xi Q+\xibar\Qbar,$
these transform as \dkmfour:\foot{In the 1-instanton sector, these
SUSY transformations are equivalent, up to an $SU(2)$ gauge transformation,
to those of Novikov, Shifman, Vainshtein and Zakharov \NSVZ.}
\eqna\susyalgebra
$$\eqalignno{
\delta a_{\alpha\dalpha}\ &=\ \xibar_{\dalpha}\M_\alpha
\qquad&\susyalgebra a \cr
\delta\M_\gamma\ &=\ -4ib\xi_{\gamma}
\qquad&\susyalgebra b \cr}$$
This algebra allows us to promote
 $a$ to a space-time-constant 
``superfield'' $a(\thetabar)$ in an
obvious way:\foot{Here we ignore the action of the 
$Q_\alpha$. 
These generators correspond to supersymmetries which are broken by the
self-dual gauge-field configuration and they 
act on the moduli in a trivial way.}
\eqn\apromote{a_{\alpha\dalpha}
\ \rightarrow\ a_{\alpha\dalpha}(\thetabar)\ =\ 
 e^{\thetabar\Qbar}\times  
a_{\alpha\dalpha}
\ =\ 
a_{\alpha\dalpha}+\thetabar_{\dalpha}\M_\alpha\ .}
In superfield language the bosonic and fermionic constraints \crucial a and
\zmcons a assemble naturally into a supermultiplet of constraints,
namely \dkmfour:
\eqn\superadhm{\abar(\thetabar) a(\thetabar)\ =\ \big(\abar(\thetabar)
 a(\thetabar)\big)^T\ \propto\ \delta^\dbeta{}_\dalpha\ .}
The scalar piece of \superadhm\ gives \crucial a and the $\CO(\thetabar)$
piece gives \zmcons a, while the $\CO(\thetabar^2)$ piece is
``auxiliary'' as it is satisfied automatically.

\subsec{Ansatz for the measure}

In this section we will discuss the
$N=1$ superinstanton measure $\dmunphys$, for arbitrary topological
number $n$. As the small-fluctuations determinants in a self-dual
background cancel between the bosonic and fermionic sectors in a
supersymmetric theory \dadda, 
the relevant measure is the one inherited from the 
Feynman path integral on changing variables from the fields 
to the collective coordinates which parametrize the instanton moduli space
${\rm M}_{\rm phys}$. In principle the super-Jacobian for this change of
variables can be calculated by evaluating the normalization matrices of
the appropriate bosonic and fermionic zero-modes. 
In practice, this involves solving the ADHM constraints (2.5) and can
only be accomplished for $n\leq 2$. 

As discussed in Section 1, we will
pursue an alternative approach to the problem of determining the
correct measure. The first step is to formally undo the $O(n)$ quotient
described in Eq.~\modspace\ and define an unidentified measure, $\dmun$, for
integration over the larger moduli space ${\rm M}$:
\eqn\dmudefI{\int_{{\rm M}_{\rm phys}}\dmunphys\ \equiv\ 
{1\over\VolOn}\,\int_{{\rm M}}\dmun}
The correctly normalized volumes for the $O(n)$ groups follow from
\eqn\Onrec{O(n)\ =\ {O(n)\over O(n-1)}\times {O(n-1)\over O(n-2)}
\times\cdots\times {O(2)\over O(1)}\times O(1)}
and
\eqn\spheredef{{O(n)\over O(n-1)}\ =\ {SO(n)\over SO(n-1)}\ =\ S^{n-1}}
where $S^{n-1}$ is the $(n-1)$-sphere. Consequently the group volumes
are fixed by the recursion relation
\eqn\volrec{{\rm Vol}\big(O(n)\big)\ =\ 
2\cdot{\rm Vol}\big(SO(n)\big)\ =\ {2\pi^{n/2}\over\Gamma(n/2)}\cdot
{\rm Vol}\big(O(n-1)\big)}
together with the initial condition
\eqn\initcond{{\rm Vol}\big(O(1)\big)\ =\ 2\ .}

We will now seek a measure, $\dmun$, with the following 
five properties:\hfil\break\indent
(i) $O(n)$ invariance; \hfil\break\indent
(ii) \susy\ invariance; \hfil\break\indent
(iii)  a net  of $8n$ unconstrained 
bosonic integrations and also $4n$ unsaturated Grassmann
integrations over the parameters of the adjoint
zero modes  ($8n$ in the $N=2$ case);\hfil\break\indent
(iv) cluster decomposition in the dilute-gas limit  of large
space-time separation between instantons; \hfil\break\indent   
(v) agreement with known formulae in the 1-instanton sector.
\hfil\break\noindent
Taken together, these are very restrictive requirements and we claim
that they uniquely determine the measure $\dmun$. In particular, we 
conclude that the following Ansatz is the unique solution to  
conditions (i), (ii) and (iii):
\def\ijn{{(ij)^{}_n}}
\def\ijtwo{{(ij)^{}_2}}
\def\ijna{{\langle ij\rangle^{}_n}}
\def\ijnone{{(ij)^{}_{n-1}}}
\def\ijnonea{{\langle ij\rangle^{}_{n-1}}}
\eqn\dmudef{\eqalign{\int\dmunphys\ &\equiv\ {1\over\VolOn}\,\int\dmun
\cr&=\ {C_n\over\VolOn}\int\prod_{i=1}^nd^4w_id^2\mu_i
\prod_{\ijn}d^4a'_{ij}d^2\M'_{ij}
\cr&\times\ \prod_{\ijna}\prod_{c=1,2,3}
\delta\big(\quarter\trtwo\tau^c[(\abar a)_{i,j}-
(\abar a)_{j,i}]\big)\,\delta^2\big((\abar \M)_{i,j}-
(\abar \M)_{j,i}\big)\ .}}
where the collective coordinates $w_i,$ $\mu_i,$ $a'_{ij}$ and $\M'_{ij}$ were
defined in Eqs.~\bcanonical\ and \Mdef. The notation
$\ijn$ and $\ijna$, which we will use 
for symmetric and antisymmetric $n\times n$
matrices respectively, stands for the ordered pairs $(i,j)$ restricted
as follows:
\eqna\ordered
$$\eqalignno{\ijn\,&:\qquad 1\le i\le j\le n &\ordered a
\cr
\ijna\,&:\qquad 1\le i< j\le n &\ordered b
\cr}$$
Further, we will show below that the overall numerical constants $C_{n}$ are
determined by condition (iv) up to a single normalization which is, in turn,
fixed by condition (v).  

The two $\delta$-functions in \dmudef\ implement the constraints
\crucial a and \zmcons a, respectively. 
In the case of the 1-instanton measure these constraints
disappear,  and one simply has
\eqn\onesimp{\int\dmuonephys\ \equiv\ \hf\int\dmuone
\ =\ \hf C_1\int d^4wd^2\mu d^4a'd^2\M'\ .} 
This is precisely
't Hooft's 1-instanton measure \tHooft, rewritten in the quaternionic
notation of ADHM. In particular the position, size, and $SU(2)$
iso-orientation of the instanton are given, respectively, by
$a',$ $|w|$, and $w/|w|,$  with the fermionic quantities
in \onesimp\ 
denoting their respective superpartners. Also
 $\hf C_1$ is 't Hooft's scheme-dependent 1-instanton factor.

In the remainder of Sec.~2 we verify that the measure \dmudef\
satisfies the above mentioned properties and then perform a highly
non-trivial check on our Ansatz by comparing it to the known results
in the two-instanton sector. 

\subsec{$O(n)$ invariance, SUSY invariance, and dimensional power counting}

The $O(n)$ invariance of this measure
is obvious by inspection. As for \susy\ invariance under
\susyalgebra{},  this too is obvious for $\xi Q$, while for $\xibar\Qbar$
the reasoning is as follows: the argument of the
second $\delta$-function in \dmudef\ (which implements \zmcons a)
is invariant, while that of the
first  $\delta$-function (which implements \crucial a)
transforms into itself plus an admixture of  the second 
under \susyalgebra a, so that the product of 
$\delta$-functions is an invariant. The underlying reason for this
 is, of course, that the constraints \crucial{} and 
\zmcons{} form a supermultiplet, Eq.~\superadhm.

Next we verify the counting requirement (iii). Recall that by the
rules of Grassmann integration,\foot{We remind the reader of the complete
``Gradshteyn and Ryzhik'' for Grassmann integration: $\int d\chi=0$
and $\int d\chi\,\chi=1$.}
 a $\delta$-function of a Grassmann-valued
argument may simply be replaced by the argument itself. So, of
the $n^2+3n$ fermionic  $\mu_i$ and $\M'_{ij}$ 
modes in
\dmudef, $n(n-1)$ of them
are saturated by the second $\delta$-function, leaving $4n$ unbroken
gaugino zero
modes as required. As a further test of our Ansatz 
one may check  that \dmudef\  has the correct bosonic dimensionality.
The bosonic sector scales like
\eqn\bosdim{[a]^{\,(2n^2+6n)-(3n^2-3n)+(n^2-n)}\ \sim\ 
[a]^{8n}\ ,}
the three terms in the exponent coming, respectively, from the $w_i$
and $a'_{ij}$
integration variables, from the first $\delta$-function, and from the
second $\delta$-function.

\subsec{Cluster decomposition}

Next we examine the clustering property (iv), which turns out to
be the least obvious, and most stringent, of the requirements.
We will analyze the limit in which one of the instanton position moduli is
far away from all the others, and demand that the measure
\dmudef\ factor approximately into a product of a 1-instanton and an
$(n-1)$-instanton measure. Recall that in the limit of large
separation, the space-time positions of the $n$ individual
instantons making up the topological-number-$n$
configuration may simply be identified with the $n$ diagonal elements
$a'_{ii}$ \CWS. A convenient clustering limit is then
\eqn\onecluster{|a'_{nn}|\ \rightarrow\ \infty}
with all other elements of $a$ remaining ``order unity'' (or smaller,
if dictated by the ADHM constraints).
Of course \onecluster\ is not an $O(n)$-invariant statement. We can rewrite
\onecluster\ as the statement that a rank-one submatrix $h$ of $a',$
defined as
\eqn\hdef{h\ =\ q\cdot V\cdot V^T\ ,}
where $q$ is a quaternion and $V$ is a unit-normalized $n$-vector in
$\bigR^n,$ becomes large: 
\eqn\qlimit{|q|\ \rightarrow\ \infty\ .}
Equation~\onecluster\ then corresponds to the choice 
\eqn\qequals{V\ =\ \pmatrix{0\cr\vdots\cr0\cr1}\ ,\qquad
q\, =\,a'_{nn}\ .}

Note that an $O(n-1)$ subgroup of $O(n)$ leaves the choice \hdef\ and
\qequals\
invariant; \hdef\ is only acted on by the coset $O(n)/O(n-1)$ which
sweeps the vector $V$ through the $(n-1)$-sphere $S^{n-1}.$  
To make this more precise, we parametrize $SO(n)\subset O(n)$ by the set of
$n\times n$ generators $t_{ij},$ $1\le i<j\le n,$ defined by their matrix
elements
\eqn\tmatdef{\big(t_{ij}\big)_{kl}\ =\ \delta_{ik}\delta_{jl}-
\delta_{il}\delta_{jk}\ .}
Every $g\in SO(n)$ can be written as \gilmore
%\eqn\ginSOn{g\ =\ \exp\Big(-\sum_{\ijna}\alpha_{ij}\,t_{ij}\Big)}
\eqn\ginSOn{g\ =\ \prod_{j=1}^n\exp\left(-\sum_{i=1}^{j-1}\alpha_{ij}\,
t_{ij}\right)\ .}
In these coordinates the properly normalized Haar measure for $SO(n)$
takes the form of a nested product of cosets, as per Eq.~\Onrec:
\eqn\Haar{dg\,=\,\prod_{j=1}^n dg_j\ ,\qquad dg_j\,=\,
\Big(\prod_{i=1}^{j-1}d\alpha_{ij}
\Big){1\over\sqrt{1-\sum_{i=1}^{j-1}\alpha_{ij}^2}}}
while the group volumes \volrec\ follow from
\eqn\volexpl{{\rm Vol}\big(SO(n)\big)\ \equiv\ 
\int_{SO(n)}dg\ =\ \prod_{j=1}^n\,2\int_{D(j-1)}dg_j}
Here the domain of integration $D(j-1)$ is the unit $(j-1)$-ball,
 and the factors of 2 count the two hemispheres of each coset.
%
%\eqn\gpvols{{\rm Vol}\big(O(n)\big)\ =\ 2\cdot
%{\rm Vol}\big(SO(n)\big)\ =\ 2\int_{\Gamma_n}\ \prod_{\ijna}d\alpha_{ij}\ .}
For infinitesimal $\alpha_{ij},$ $g$ acts on the submatrix $h$ defined
by Eqs.~\hdef\ and \qequals\ as follows:
\eqn\gacts{\eqalign{a'_{nn}\cdot
\pmatrix{0\cr\vdots\cr0\cr1}
\matrix{\big(0,\cdots,0,1\big)\cr{}\cr{}\cr{}}
\ &\longrightarrow\
a'_{nn}\,g^T\cdot
\pmatrix{0\cr\vdots\cr0\cr1}
\matrix{\big(0,\cdots,0,1\big)\cr{}\cr{}\cr{}}
\cdot g
\cr
&=\ a'_{nn}\cdot\pmatrix{0&\cdots&0&\alpha_{1n}\cr
\vdots&{}&\vdots&\vdots\cr
0&\cdots&0&\alpha_{n-1,n}\cr
\alpha_{1n}&\cdots&\alpha_{n-1,n}&1\cr}\ +\ \CO(\alpha_{ij}^2)\ .}}
Equation \gacts\ defines the infinitesimal
action of the coset $O(n)/O(n-1)\cong
SO(n)/SO(n-1)$ on $h$.

We can now state precisely what is meant by the clustering condition in
the limit $|a'_{nn}|\rightarrow\infty.$
It is important to stress that clustering is a property of the
unidentified measure $\dmun,$ rather than of the physical measure
$\dmunphys$ in which points on the $O(n)$ orbit are identified.
This distinction
 is emphasized in Refs.~\refs{\GMO,\Osborn} (albeit in a formalism
in which it is a finite subgroup of $O(n)$ rather than the full $O(n)$
that is being modded out). Furthermore, 
for the unidentified measure, one cannot
simply demand $\dmun\,\rightarrow\,\dmunone\times\dmuone$ since the number
of bosonic differentials on the left-hand side exceeds that of the
right-hand side by $\dim(O(n))-\dim(O(n-1))=n-1.$ Instead, the
proper clustering condition reads
\eqn\demand{\dmun\ \buildrel{|a'_{nn}|\rightarrow\infty}\over
\longrightarrow\ 
\dmunone\times\dmuone\times dS^{n-1}}
Here $\dmunone$ is built from the variables
$\{w_j,\mu_j,a'_{ij},\M'_{ij}\}$ with $1\le i
\le j\le n-1$; $\dmuone$
is built from  $\{w_n,\mu_n,a'_{nn},\M'_{nn}\}$; and
\eqn\dSdef{dS^{n-1}\ =\ \prod_{i=1}^{n-1}d\alpha_{in}\ }
 in the notation of Eqs.~\ginSOn-\gacts. Note from Eq.~\Haar\
that Eq.~\dSdef\ is
only correct for $V^T$ in an infinitesimal neighborhood of $(0,\cdots,0,1)$,
i.e. for infinitesimal $\alpha_{in}$;
this is all that is actually needed for present purposes.

 The calculation below  proceeds as follows.
In the fermionic sector, 
the $n-1$ extra \hbox{$\delta$-functions} on the left-hand side
of \demand, namely 
$\prod_{i=1}^{n-1}\delta^2\big((\abar\M)_{i,n}-(\abar\M)_{n,i}\big),$
which have no counterparts on the right-hand side,
are killed upon integration of the $n-1$ extra differentials
$\prod_{i=1}^{n-1}d^2\M'_{in}$.
The bosonic sector is more complicated: the extra
\hbox{$\delta$-functions}
$\prod_{i=1}^{n-1}\prod_{c=1,2,3}\,
\delta\big(\quarter\trtwo\tau^c[(\abar a)_{i,n}-
(\abar a)_{n,i}]\big)$
are killed by three-quarters of the extra bosonic differentials
$\prod_{i=1}^{n-1}\int d^4a'_{in}$; the remaining one-quarter 
assemble to form
 the differential $dS^{n-1}$ on the right-hand side of Eq.~\demand.
In this way we shall verify the condition \demand, and in so doing,
obtain a formula for the overall factors $C_n.$
 Here are the details:

Let us return to the measure $\dmun,$ Eq.~\dmudef, and perform, first, the
subset of Grassmann integrations
\eqn\subsetf{\int\prod_{i=1}^{n-1}d^2\M'_{in}\,
\delta^2\big((\abar\M)_{i,n}-(\abar\M)_{n,i}\big)\times\cdots}
The dots represent the remaining $\delta$-functions in the integrand
of \dmudef. We expand
\eqn\expandf{(\abar\M)_{i,n}-(\abar\M)_{n,i}\ =\ -\abar'_{nn}\M'_{in}
+\sum_{k=1}^{n-1}\abar'_{ik}\M'_{kn}+\cdots,}
 the dots representing modes other than the $\M'_{kn}$ modes. Hence the
integration \subsetf\ kills the $\delta$-functions and produces the Jacobian
\eqn\Jacfdef{J_\M\ =\ |a'_{nn}|^{2(n-1)}\ +\ \cdots ,}
neglecting subleading powers of $|a'_{nn}|.$
Next, one carries out the analogous subset of bosonic integrations
\eqn\subsetb{\int\prod_{i=1}^{n-1}d^4a'_{in}\,\prod_{c=1,2,3}\,
\delta\big(\quarter\trtwo\tau^c[(\abar a)_{i,n}-
(\abar a)_{n,i}]\big)\times\cdots}
in the following manner. First one changes quaternionic variables to

\def\ahat{{\hat a}}
\eqn\hatadef{a'_{in\,\alpha\dalpha}
\ =\ a'_{nn\,\alpha\dbeta}\,\ahat^{\dbeta}_{in\,\dalpha}\ 
,\qquad 1\le i\le n-1\ ,}
where $\ahat_{in}$ 
in turn is divided into self-bar (SB) and anti-self-bar (ASB)
pieces, thus:
\def\asb{\hat a^{\sst\rm SB}}
\def\aasb{\hat a^{\sst\rm ASB}}
\eqn\sbdef{\ahat_{in}\, =\, \asb_{in}+\aasb_{in}\ ,\quad
\bar{\hat a}^{\sst\rm SB}_{in}\,=\,
{\hat a}^{\sst\rm SB}_{in} \ ,\quad
\bar{\hat a}^{\sst\rm ASB}_{in}\,=\,
-{\hat a}^{\sst\rm ASB}_{in} \ .}
Note that ${\hat a}^{\sst\rm SB}_{in}$
is a scalar: $({\hat a}^{\sst\rm SB}_{in})^\dbeta{}_\dalpha\propto
\delta^\dbeta{}_\dalpha.$
In these variables the integration reads
\eqn\measurehat{\int
\prod_{i=1}^{n-1}d^4a'_{in}\ =\ |a'_{nn}|^{4(n-1)}\int\prod_{i=1}^{n-1}
d^3\aasb_{in}\,d\asb_{in}\ .}
and the argument of the $\delta$-function in \subsetb\ becomes
\eqn\expandb{\eqalign{&(\abar a)_{i,n}-(\abar a)_{n,i}\cr&=\ 
-2\aasb_{in}\,|a'_{nn}|^2+\sum_{k=1}^{n-1}\big(\abar'_{ik}a'_{nn}(\asb_{kn}
+\aasb_{kn})-(\asb_{kn}-\aasb_{kn})\abar'_{nn}a'_{ki}\big)\ +\cdots}}
where the dots represent terms that do not depend on the $\ahat_{kn}.$
Performing the $d^3\aasb_{in}$ integration kills the $\delta$-functions
in \subsetb\ and produces the Jacobian 
\eqn\Jadhmdef{J_{a}\ =\ \left({1\over|a'_{nn}|^2}\right)^{3(n-1)}\
+\ \cdots,}
again neglecting terms subleading in $|a'_{nn}|$. Note that the powers
of $|a'_{nn}|$ precisely cancel among Eqs.~\Jacfdef, \measurehat\ and
\Jadhmdef. 

It remains to carry out the $n-1$  integrations over
the $\asb_{in}$. But these integration variables, viewed as 
infinitesimals,
are precisely the generators \gacts\ of  $O(n)/O(n-1)$ that sweep the vector
$V$ in \hdef\ through  $S^{n-1}$:
\eqn\sweeper{{\hat a}^{\sst\rm SB}_{in}\ =\ \alpha_{in}\ ,\qquad
\prod_{i=1}^{n-1}d{\hat a}^{\sst\rm SB}_{in}\ =\ dS^{n-1}\ .}

\def\atilde{{\tilde a}}
\def\Mtilde{{\tilde \M}}
Gathering the results to this point, we have shown that in the limit
$|a'_{nn}|\rightarrow\infty,$
\eqn\mugoes{\eqalign{\int\dmun\quad & 
\longrightarrow\quad {C_n}\int dS^{n-1}\times
\int dw_nd\mu_nd^4a'_{nn}d^2\M'_{nn}\cr&\times
\int\prod_{i=1}^{n-1}d^4w_id^2\mu_i
\prod_{\ijnone}d^4a'_{ij}d^2\M'_{ij}
\cr&\times\ \prod_{\ijnonea}\prod_{c=1,2,3}
\delta\big(\quarter\trtwo\tau^c[(\abar a)_{i,j}-
(\abar a)_{j,i}]\big)\,\delta^2\big((\abar \M)_{i,j}-
(\abar \M)_{j,i}\big)\ .}}
The second integral in the
first line is proportional to the 1-instanton measure $\dmuone$
as anticipated.
Let us define $\atilde$ and $\Mtilde$ to be the truncated versions of $a$ and
$\M$, with the last row and column lopped off:
\eqn\atildedef{\atilde\ =\ \pmatrix{w_1&\ldots&w_{n-1}\cr
a'_{11}&\ldots&a'_{1,n-1}\cr\vdots&\ddots&\vdots\cr
a'_{1,n-1}&\ldots&a'_{n-1,n-1}\cr}
\ ,\qquad
\Mtilde\ =\ \pmatrix{\mu_1&\ldots&\mu_{n-1}\cr
\M'_{11}&\ldots&\M'_{1,n-1}\cr\vdots&\ddots&\vdots\cr
\M'_{1,n-1}&\ldots&\M'_{n-1,n-1}\cr}
}
We still need to verify that the last two lines of \mugoes\ are proportional
to the $(n-1)$-instanton measure $\dmunone,$ meaning that $a$ and $\M$
can be approximately replaced by $\atilde$ and $\Mtilde$ inside the
$\delta$-functions. The
 error in making these approximations is:
\def\atildebar{\bar{\tilde a}}
\eqn\diffb{[
(\abar a)_{i,j}-(\abar a)_{j,i}]-[(\atildebar\atilde)_{i,j}-(\atildebar
\atilde)_{j,i}]\ =\ |a'_{nn}|^2\,\big[\,\aasb_{jn}\,,\,\aasb_{in}\,
\big]}
and
\eqn\difff{[(\abar \M)_{i,j}-(\abar \M)_{j,i}]-
[(\atildebar \Mtilde)_{i,j}-(\atildebar \Mtilde)_{j,i}]\ =\
\aasb_{jn}\abar'_{nn}\M'_{in}-\aasb_{in}\abar'_{nn}\M'_{jn}\ ,}
where we neglect the $\asb_{in}$ as we are focusing on an
infinitesimal neighborhood of $V$.
{}From Eqs.~\expandb\ and \expandf\ one learns that
\eqn\scalelike{\aasb_{in}\ \sim\ |a'_{nn}|^{-2}\ ,\qquad
\M'_{in} \ \sim\ |a'_{nn}|^{-1}}
so that the right-hand sides of \diffb\ and \difff\ indeed vanish like
$|a'_{nn}|^{-2}$ in the clustering limit, as desired.
The clustering condition \demand, applied to Eq.~\mugoes,
 then collapses to the simple numerical
recursion
\eqn\recrel{C_n\ =\ C_{n-1}\cdot C_1}
or equivalently
\eqn\Cnsolve{C_n\ =\ (C_1)^n\ .}
This formula completes the specification of the $N=1$ \susic\ instanton
measure \dmudef.

\subsec{Agreement in the 2-instanton sector}

Finally we verify  that our Ansatz \dmudef\ and
\Cnsolve\ for the
$N=1$ \susic\ measure $\dmunphys$ agrees with previously known results
for topological number $n=1$ \tHooft\ and $n=2$ \refs{\Osborn,\dkmone}.
The case $n=1$ was already discussed in Eq.~\onesimp\ \it ff\rm. The case
$n=2$ is more interesting. Following \refs{\Osborn,\dkmone},
$\dmutwophys$ is known from first principles to have the following form:
\def\Stwo{{\cal S}_2}
\def\Jbose{J_{\rm bose}}
\def\Jfermi{J_{\rm fermi}}
\eqn\dmuOsb{\int\dmutwophys\ =\ {1\over\Stwo}\int d^4w_1d^4w_2
d^4a'_{11}d^4a'_{22}d^2\mu_1d^2\mu_2d^2\M'_{11}d^2\M'_{22}
\,\big(\,J_\bose/J_\fermi\,\big)^{1/2}
}
The definition of the 
zero-mode Jacobians $\Jbose$ and $\Jfermi$ and the discrete
group-theoretic $\Stwo$ will be reviewed below.
Note that this particular form for the measure  breaks $O(2)$
invariance as it involves integration over only the diagonal elements
of $a'$ and $\M',$ the off-diagonal elements having been eliminated
by the explicit resolution of all the
 constraints (which is elementary to do only for $n=2$). 
At the same time it breaks SUSY invariance,
since (as reviewed shortly)  $\Jbose$ and $\Jfermi$
are purely bosonic expressions with no fermion bilinear parts.
Our goal is to demonstrate that this measure is nevertheless equivalent
to the $O(2)$- and SUSY-invariant form \dmudef. In fact, the known
expression \dmuOsb\ is precisely an ``$O(2)$-gauge-fixed'' version of
\dmudef. (We put this in quotes because $O(2)$ invariance
 is purely internal to the ADHM
construction and has nothing to do with the usual choice of
space-time gauge.) As shown below, to recover an $O(2)$-invariant form we
will integrate \dmuOsb\ over $O(2)$ orbits; pleasingly,
SUSY invariance is recovered simultaneously. 

The quantities $\Jfermi$ and $\Jbose$ that enter \dmuOsb\ were obtained
in Refs.~\dkmone\ and \Osborn, respectively. Up to a multiplicative constant,
$\Jfermi$ has the form\foot{This expression
 is the square-root of the formula in \dkmone,
as the $N=1$ theory contains half as many adjoint fermion zero modes
as the $N=2$ theory.}
\eqn\Jfermidef{\Jfermi^{1/2}\ \propto\ {H\over|a_3|^2}}
where
\eqn\Hdef{H\ =\ |w_1|^2+|w_2|^2+4|a'_{12}|^2+4|a_3|^2}
and $a_3$ (likewise $a_0$, needed below) 
is shorthand for the linear combination
\eqn\athreedef{a_3\ =\ \hf(a'_{11}-a'_{22})\ ,\quad
a_0\ =\ \hf(a'_{11}+a'_{22})\ .}

\def\Sigphi{\Sigma^\phi}
\def\Rphi{R_\phi}
The construction of $\Jbose$ is considerably more intricate, due to the
nonlinearity of the bosonic constraint \crucial a and the need to
enforce the background gauge condition, and is the principal
 achievement of \Osborn. For $n=2$ the general solution
to \crucial a is easily obtained, and reads:
\eqn\aonedef{a'_{12}\ =\ {1\over4|a_3|^2}\,a_3(\wbar_2w_1-\wbar_1w_2+
\Sigma)\ .}
Here $\Sigma^\dalpha{}_\dbeta$ is an arbitrary scalar-valued function
of $\{a_3,a_0,w_1,w_2\}$ (which as per \dmuOsb\ we select to be our set of
$8n=16$ independent bosonic variables):
\eqn\Sigmadef{\Sigma^\dalpha{}_\dbeta\ =\ \delta^\dalpha{}_\dbeta
\,\Sigma(a_3,a_0,w_1,w_2)\ .}
{}From \aonedef\ we can solve instead for $\Sigma\,$:
\eqn\Sigsolve{\Sigma\ =\ 2(\abar^{}_3a'_{12}+\abar'_{12}a^{}_3)\ .}
In these coordinates, Osborn's result for the bosonic Jacobian reads
\Osborn
\eqn\Jbosepropto{J_\bose^{1/2}\ \propto\ {H\over|a_3|^4}\,
\Big|\,|a_3|^2
\,-\,|a'_{12}|^2\,-\,\eighth{d\Sigma^\phi
\over d\phi}\big|_{\phi=0}\,\Big|\ ,}
so that 
\eqn\Jratio{\big(\,J_\bose/J_\fermi\,\big)^{1/2}\ =\ C_2\,
{\Big|\,|a_3|^2
\,-\,|a'_{12}|^2\,-\,\eighth{d\Sigma^\phi
\over d\phi}\big|_{\phi=0}\,\Big|\over
|a_3|^2}\ .}
The quantity 
$\Sigphi$ is defined as follows. The angle $\phi$ parametrizes
the $SO(2)\subset O(2)$ transformation 
\eqn\Rphidef{\Rphi\ =\ 
\pmatrix{  \cos\phi & \sin\phi \cr
-\sin\phi & \cos\phi }\ ,\qquad0\le\phi<2\pi\ .}
Under the $SO(2)$ transformation \trans, the bosonic coordinates transform
as
\eqna\bostrans
$$\eqalignno{(w_1,w_2)\ &\rightarrow\ (w_1^\phi,w_2^\phi)\ \equiv\
(w_1,w_2)\cdot R_\phi &\bostrans a\cr
(a^{}_3,a'_{12})\ &\rightarrow\ (a_3^\phi,a_{12}^{\prime\phi})\ \equiv\
(a^{}_3,a'_{12})\cdot R_{2\phi} &\bostrans b\cr
a_0\ &\rightarrow\ a_0^\phi\ \equiv\ a_0 &\bostrans c\cr
}$$
and likewise for their corresponding fermionic superpartners in $\M$. 
Consequently\foot{This is an interesting self-referential equation
since $a_3^\phi$ depends on $a'_{12}$ which, in turn, depends on $\Sigma.$}
\eqn\Siggoes{\Sigma(a_3,a_0,w_1,w_2)\ \rightarrow\ \Sigma^\phi\ \equiv\
\Sigma(a_3^\phi,a^{}_0,w_1^\phi,w_2^\phi)\ .}

Since $\Sigma$ transforms nontrivially under $O(2)$, an ``$O(2)$-gauge-fixing''
prescription is locally equivalent to a particular functional choice
for $\Sigma,$ but the \it global \rm equivalence is not guaranteed
\Osborn:
If, on the one hand, the $O(2)$ redundancy group
is broken completely, then each point
on the ADHM manifold is indeed in 1-to-1 correspondence with a physical
2-instanton configuration. 
On the other hand, it generically happens that specifying $\Sigma$ does
not break $O(2)$ completely, but leaves a residual \it discrete \rm subgroup
$G_2\subset O(2)$. The action of $G_2$ must then be modded out in
constructing the physical measure $\dmutwophys$; this is precisely the
overall factor \Osborn
\eqn\Stwodef{\Stwo \ =\ \dim{G_2}}
that appears in  Eq.~\dmuOsb. Intuitively, $G_2$ will include the permutation
group $P(2)$ which exchanges the identities of the two instantons.
In practice, $G_2$ is often much bigger than $P(2)$.
For instance, for the
simple choice $\Sigma\equiv0,$ $G_2$ is the dihedral group $D_8$, and
consequently $\Stwo=\dim(D_8)=16$ \dkmone; other discrete groups can occur
for different choices of $\Sigma$.

The remaining quantity introduced above that needs 
to be specified  is the parameter $C_2$ in Eq.~\Jratio; in our conventions
it contains all collective-coordinate-independent multiplicative factors
in \dmuOsb\ apart from $\Stwo^{-1}$.
In fact $C_2$ is the same parameter as entered the $O(2)$-invariant
measure, \dmudef\ and \Cnsolve. To see this, it suffices to consider the limit
 $|a'_{22}|\rightarrow\infty\,$. For nonpathological $\Sigma$ this
means 
\eqn\JJgoes{{\Big|\,|a_3|^2
\,-\,|a'_{12}|^2\,-\,\eighth{d\Sigma^\phi
\over d\phi}\big|_{\phi=0}\,\Big|\over
|a_3|^2}\ \longrightarrow\ 1}
so that the clustering property forces
\eqn\asbefore{C_2\ =\ (C_1)^2}
as per Eq.~\Cnsolve\ above.

Now let us, by hand, restore 
$O(2)$ invariance to the $O(2)$-gauge-fixed measure
\dmuOsb. The first step is to ``integrate in'' the off-diagonal
elements $a'_{12}$ and $\M'_{12}$ by inserting the factors of unity
\eqn\unityb{1\, =\, 16|a_3|^4\int d^4a'_{12}\,
\prod_{c=1,2,3}
\delta\big(\quarter\trtwo\tau^c[(\abar a)_{1,2}-(\abar a)_{2,1}]\big)
\delta\big(\abar_3^{}a'_{12}+\bar a'_{12}a^{}_3-\hf
\Sigma(a_3,a_0,w_1,w_2)\big)}
and
\eqn\unityf{1\ =\ {1\over4|a_3|^2}\int d^2\M'_{12}\,
\delta^2\big((\abar\M)_{1,2}-(\abar\M)_{2,1}\big)}
The second $\delta$-function in \unityb\ enforces the ``$O(n)$-gauge-fixing''
\Sigsolve; the other two \hbox{$\delta$-functions} in \unityb-\unityf\
are the by-now-familiar
 implementations of the constraints \crucial a and \zmcons a.
Next one performs a change of dummy integration variables $a\rightarrow
a^\phi$ and $\M\rightarrow\M^\phi$ as defined by Eqs.~\bostrans{}. 
The $\delta$-function in \unityf, and the first
$\delta$-function in \unityb, are unchanged, since their arguments are
formed from the $O(2)$-invariant matrix
$$\pmatrix{0&1\cr-1&0}\ .$$
Likewise $H$ is an invariant. The only non-invariants are the second
$\delta$-function in \unityb, and the piece
$\big|\,|a_3|^2\,-\,|a'_{12}|^2\,-\,\eighth(d\Sigma^\phi/
d\phi)|_{\phi=0}\,\big|$ from $\Jbose^{1/2}$.
Together they transform nontrivially, as follows:
\eqn\rotating{\eqalign{
\Big|\,|a_3|^2\,-\,|a'_{12}|^2\,-\,\eighth{d\Sigma^\phi\over 
d\phi}\big|_{\phi=0}\,\Big|\,\cdot\,&
\delta\big(\abar_3^{}a'_{12}+\bar a'_{12}a^{}_3-\hf
\Sigma(a_3,a_0,w_1,w_2)\big)\cr &\longrightarrow\ 
\quarter\big|Z'(\phi)\big|\,\delta\big(Z(\phi)\big)}}
where
\eqn\Zdef{Z(\phi)\ =\ 
\abar_3^{\phi}a^{\prime\phi}_{12}+\bar a^{\prime\phi}_{12}a^{\phi}_3-\hf
\Sigma^\phi\ }
in the notation of Eq.~\bostrans{}.
Inserting another factor of unity, namely
\eqn\anotherone{1\ =\ {1\over2\pi}\int_0^{2\pi}d\phi}
and integrating over the right-hand side of \rotating\ gives a factor of
\eqn\newfactor{{1\over2\pi}\cdot{1\over4}\cdot k}
where the integer $k$ counts the  number of times in the interval $[0,2\pi]$
that $Z(\phi)$ vanishes. Obviously $k$ is related to the existence of
the discrete subgroup $G_2$ discussed above; in fact one typically 
has
\eqn\qrelation{k\ = \ \hf\Stwo}
where the factor of $\hf$ is due to the fact that in Eqs.~\anotherone\
and \Rphidef\ we are only integrating
over half the $O(2)$ group, namely $SO(2)$.\foot{The relation \qrelation\
may fail if
$\Sigma$ is taken to depend on odd powers of the $w_i$ (breaking
the ``parity'' symmetry $w_i\rightarrow-w_i$). If so, then rather
than Eq.~\anotherone\ one should integrate over both halves of the
$O(2)$ group separately and divide the sum
by $4\pi$. The final answer for $\dmutwophys$ is then the same.}

Combining Eqs.~\dmuOsb, \Jratio, \asbefore-\unityf, and
\newfactor-\qrelation\ gives, finally,
\eqn\dmuagain{\eqalign{\int\dmutwophys\ 
&=\ {(C_1)^2\over4\pi}\int\prod_{i=1}^2d^4w_id^2\mu_i
\prod_{\ijtwo}d^4a'_{ij}d^2\M'_{ij}
\cr&\times\ \prod_{c=1,2,3}
\delta\big(\quarter\trtwo\tau^c[(\abar a)_{1,2}-
(\abar a)_{2,1}]\big)\,\delta^2\big((\abar \M)_{1,2}-
(\abar \M)_{2,1}\big)\ .}}
Notice that the $\Sigma$-dependent discrete group factor $\Stwo$ introduced
in Eq.~\dmuOsb\ has
canceled out. In fact, since $4\pi={\rm Vol}(O(2)),$ we have recaptured
precise agreement with the earlier $O(2)$- and SUSY-invariant
expression \dmudef\ and \Cnsolve, as promised.  This concludes our list
of checks of the proposed $N=1$ measure \dmudef\ and \Cnsolve.

\newsec{The $N=2$ \susic\ collective coordinate integration measure}
\subsec{ADHM and $N=2$ SUSY review}

Next we turn to the $N=2$ case. (The extension to
$N=4$ is more intricate still and will be discussed in a separate
publication.) The particle content of $N=2$ \susic\ Yang-Mills theory
comprises, in addition to the gauge field $v_m$ and gaugino $\lambda_\alpha,$
an adjoint complex Higgs $A$ and Higgsino $\psi_\alpha.$ The fermion zero
modes of $\psi_\alpha$ are defined in analogy with Eq.~\lambdazm:
\eqn\psizm{(\psi_\alpha)^\dbeta{}_\dgamma\ =\ 
\Ubar^{\dbeta\gamma}\N_\gamma f\,\bbar\, U_{\alpha\dgamma}\ -\
\Ubar^\dbeta{}_\alpha \,bf\N^{\gamma T}U_{\gamma\dgamma}\ ,}
where the Weyl-spinor-valued matrix
\eqn\Ndef{\N^\gamma\ =\ \pmatrix{\nu_1^\gamma&\cdots&\nu_n^\gamma
\cr{}&{}&{}\cr
{}&\N^{\prime\gamma}&{}\cr{}&{}&{}}}
satisfies linear constraints analogous to Eq.~\zmcons{}:
\eqna\zmconsN
$$\eqalignno{\abar^{\dalpha\gamma}\N_\gamma\ &=\ -\N^{\gamma T}a_\gamma
{}^\dalpha\ ,&\zmconsN a
\cr
\N_\gamma^{\prime T}\ &=\ \N'_\gamma\ .&\zmconsN b}$$
Under $O(n)$, $\N$ transforms like $\M$ and $\Delta(x)$; see Eq.~\trans.

The construction of the classical Higgs $A$ is more involved, and
goes as follows \dkmone. $A$ has the additive form 
$A=\Aone+\Atwo$, where
\eqn\Aonedef{i\,\Aone^\dalpha_{\ \dbeta}\ =\
{1\over2\sqrtwo}\,\Ubar^{\dalpha\alpha}\big(\N_\alpha f\M^{\beta T}
-\M_\alpha f\N^{\beta T}\big)U_{\beta\dbeta}\ ,}
and 
$i\,\Atwo^\dalpha_{\ \dbeta} =
\Ubar^{\dalpha\alpha}\,\A_\alpha^{\ \beta}\,U_{\beta\dbeta} \ ,$
with $\A$  a block-diagonal constant matrix,
\eqn\blockdiag{\A_\alpha^{\ \beta}\ =\
\pmatrix{\vhiggsa_\alpha^{\ \beta}&0&\cdots&0 \cr 0&{}&{}&{}\cr
\vdots&{}&\Atot\,\delta_\alpha^{\ \beta}&{}\cr 0&{}&{}&{}}\ .}
$\vhiggsa$ is related in a trivial way to the adjoint complex
VEV $\vhiggs$ (which we point in
the $\tau^3$ direction),
\def\barvhiggsa{\bar{\cal A}_{\sst00}}
\eqn\vevbcagain{{\vhiggsa}_\alpha{}^\beta
\ =\ \textstyle{i\over2}\,\vhiggs\,\tau^3{}_\alpha{}^\beta\ ,\qquad
\barvhiggsa{}_\alpha{}^\beta
\ =\ -\textstyle{i\over2}\,\vbarhiggs\,\tau^3{}_\alpha{}^\beta\ ,
}
The $n\times n$ antisymmetric 
matrix $\Atot$ is 
defined as the
solutions to the inhomogeneous linear  equation
\eqn\thirtysomething{\bigL\cdot\Atot\ =\ \Lambda+\Lambda_f\ ,}
where $\Lambda$ and $\Lambda_f$ are the $n\times n$ antisymmetric matrices
\def\Lambdabar{\bar\Lambda}
\eqn\Lambdabardef{\Lambda_{lk}\ =\ \wbar_l\vhiggsa w_k
-\wbar_k\vhiggsa w_l\ }
and
\eqn\newmatdef{ \Lambda_f\ =\ {1\over2\sqrtwo}\,
\big(\,\M^{\beta T}\N_\beta-\N^{\beta T}\M_\beta\,\big)\ . }
$\bigL$ is a linear operator that maps the space of $n\times n$ 
 scalar-valued antisymmetric matrices onto itself. Explicitly,
if $\Omega$ is such a matrix, then $\bigL$ is defined as \dkmone
\eqn\bigLreally{\bigL\cdot\Omega\ =\ 
\hf\{\,\Omega\,,\,W\,\}\,-\,\hf\trtwo\big(
[\,\abar'\,,\,\Omega\,]a'-\abar'[\,a'\,,\,\Omega\,]\big)}
where 
$W$ is the symmetric scalar-valued $n\times n$ matrix
\eqn\Wdef{W_{kl}\ =\ \wbar_kw_l+\wbar_lw_k}
{}From Eqs.~\thirtysomething-\Wdef\ one sees that $\Atot$
transforms in the adjoint representation of $O(n)$ (i.e., like
$a',$ $\M'$ and $\N'$). 

As shown in \dkmone, 
defined in this way, the Higgs field $A$ correctly
satisfies the classical Euler-Lagrange
equation
\eqn\Higgseq{\D^2 A\ =\ \sqrtwo i\,[\,\lambda\,,\psi\,]}
where $\D^2$ is the covariant Klein-Gordon operator in the
multi-instanton background, and $\lambda$ and $\psi$ are given
by \lambdazm\ and \psizm, respectively.

\eqna\susyalgebratwo
As in the $N=1$ case, the $N=2$ \susy\ algebra may be realized directly
on these collective coordinates. Under the action of $\sum_{i=1,2}
\,\xi_iQ_i+\xibar_i\Qbar_i\,$ one has \dkmfour:\foot{As 
noted in \dkmfour, actually
the $N=2$ algebra is not faithfully represented by Eqs.~\susyalgebratwo{}. 
For instance,
the anticommutator $\{\Qonebar,\Qtwobar\}$, rather than vanishing when
acting on $a,$ $\M$ or $\N$, gives a residual $O(n)$ symmetry transformation
of the form \trans. (This is analogous to naive realizations of \susy\
that fail to commute with Wess-Zumino gauge fixing, for example.)
For present purposes this poses no problem,
as  we are always ultimately concerned with $O(n)$ singlets;
otherwise one would have to covariantize the \susy\ transformations
with respect to $O(n)$ in the standard way. This is also the reason why
Eq.~(3.15) below is not symmetric in $\thetabar_1\rightleftharpoons
\thetabar_2$.
}
$$\eqalignno{
\delta a_{\alpha\dalpha}\ &=\ \xibar_{1\dalpha}\M_\alpha
+\xibar_{2\dalpha}\N_\alpha \qquad&\susyalgebratwo a \cr
\delta\M_\gamma\ &=\ -4ib\xi_{1\gamma}-2\sqrtwo\C_{\gamma\dalpha}
\xibar_2^\dalpha  \qquad&\susyalgebratwo b \cr
\delta\N_\gamma\ &=\ -4ib\xi_{2\gamma}+2\sqrtwo \C_{\gamma\dalpha}
\xibar_1^\dalpha \qquad&\susyalgebratwo c \cr
\delta\Atot\ &=\ 0\ \qquad&\susyalgebratwo d \cr
}$$
Here $\C_{\gamma\dalpha}$ is the $(n+1)\times n$ quaternion-valued matrix
\eqn\Cdef{\C\ =\
\pmatrix{\vhiggsa w_1-w_k\Atot{}_{k1}&\cdots&\vhiggsa  
w_n-w_k\Atot{}_{kn}
\cr
{}&{}&{}\cr
{}&\big[\,\Atot\,,\,a'\,]&{}\cr
{}&{}&{}  }\ .}
Of course $\Atot$ is not an independent collective coordinate since the
invertible linear equation \thirtysomething\ fixes it in terms of
$a,$ $\M$, $\N$, and the VEV $\vhiggs$; nevertheless we will find that
the formalism simplifies when $\Atot$ is treated as though it were independent.

\def\Cf{{\cal C}_{\sst\cal N}}
As in the $N=1$ case, it is illuminating to promote $a$ to a
space-time-constant $N=2$
superfield $a(\thetabar_1,\thetabar_2)\,$ \dkmfour:
\eqn\apromote{\eqalign{a_{\alpha\dalpha}
\ &\longrightarrow\ a_{\alpha\dalpha}(\thetabar_i)\ =\ 
e^{\thetabar_2\Qbar_2}\times e^{\thetabar_1\Qbar_1}\times  
a_{\alpha\dalpha}
\cr&=\ 
a_{\alpha\dalpha}+\thetabar_{1\dalpha}\M_\alpha
+\thetabar_{2\dalpha}\N_\alpha
+2\sqrtwo\C_{\alpha\dbeta}\thetabar_2^\dbeta\thetabar_{1\dalpha}^{}
+\sqrtwo\thetabar_{1\dalpha}\thetabar_2^{\,2}\Cf{}_\alpha}}
where the Grassmann matrix $\Cf$ is defined in analogy with $\C$,
\eqn\Cfdef{\Cf\ =\
\pmatrix{\vhiggsa \nu_1-\nu_k\Atot{}_{k1}&\cdots&\vhiggsa 
\nu_n-\nu_k\Atot{}_{kn}
\cr
{}&{}&{}\cr
{}&\big[\,\Atot\,,\,\N'\,]&{}\cr
{}&{}&{}  }\ .}
The $N=2$ \susic\ ADHM constraint then reads \dkmfour:
\eqn\Superadhm{\abar(\thetabar_i) a(\thetabar_i)\ =\ \big(\abar(\thetabar_i)
 a(\thetabar_i)\big)^T\ \propto\ \delta^\dbeta{}_\dalpha\ .}
Indeed the constant component of \Superadhm\ is the bosonic constraint
 \crucial a while
the $\thetabar_1$ and $\thetabar_2$ components are the fermionic constraints
\zmcons a and
\zmconsN a, respectively. The $\thetabar^{}_{1\dalpha}\thetabar{}_{2\dbeta}$
component of \Superadhm\ gives
\eqn\newcolor{\Lambda_f\  =\ \Cbar a-(\Cbar a)^T\ ,}
which appears to be new, but is actually  
just a concise rewriting of Eq.~\thirtysomething. 
The remaining $\thetabar_i$ components of \Superadhm\ turn out to 
be ``auxiliary'' as they contain
no new information. Some are satisfied trivially, while others
boil down to linear combinations of the previous relations.

\subsec{Ansatz for the $N=2$ measure}

For arbitrary topological number $n$ 
we can now write down what we assert to be
the unique $N=2$ \susic\ collective coordinate measure
which respects criteria (i)-(iii)
listed in Sec.~2.2 above:
\eqn\dmudeftwo{\eqalign{\int\dmunphys\ &\equiv\ {1\over\VolOn}\,\int\dmun
\cr&=\ {C'_n\over\VolOn}\int\prod_{i=1}^nd^4w_id^2\mu_i d^2\nu_i
\prod_{\ijn}d^4a'_{ij}d^2\M'_{ij}d^2\N'_{ij}
\prod_{\ijna}d(\Atot)_{i,j}
\cr&\times\ \prod_{\ijna}\prod_{c=1,2,3}
\delta\big(\quarter\trtwo\tau^c[(\abar a)_{i,j}-
(\abar a)_{j,i}]\big)\,\delta^2\big((\abar \M)_{i,j}-(\abar \M)_{j,i}\big)
\cr&\qquad\qquad\qquad
\times\ \delta^2\big((\abar \N)_{i,j}-(\abar \N)_{j,i}\big)\,
\delta\big(\,(\bigL\cdot\Atot-\Lambda
-\Lambda_f)_{i,j}\,\big)\ 
 .}}
This expression appears to depend on the VEV
$\vhiggs$ which enters the definition of $\Lambda$, but this dependence is
fictitious, since the $\Atot$ integration just gives $\det^{-1}\bigL$
independent of $\vhiggs.$ Thus we could instead rewrite the final
$\delta$-function simply as $\delta((\bigL\cdot\Atot)_{i,j})$ but
we prefer to maintain a parallel treatment of the spin-0, spin-$\hf$ and
spin-1 constraints \thirtysomething, \zmcons a and \crucial a. It is
for this reason that we treat $\Atot$ as an independent collective
coordinate in \dmudeftwo\ rather than eliminate it using \thirtysomething.
We will solve for the constants $C_n'$ in Sec.~3.4 below, using the
clustering criterion (iv) as before.

In the 1-instanton sector all these constraints disappear, and
Eq.~\dmudeftwo\ collapses to the known
't Hooft measure \tHooft
\eqn\onesimptwo{\int\dmuonephys\ \equiv\ \hf\int\dmuone
\ =\ \hf C'_1\int d^4wd^2\mu d^2\nu d^4a'd^2\M'd^2\N'\ ,}
where $\hf C'_1$ is the appropriate scheme-dependent multiplicative factor.

\subsec{$O(n)$ invariance, $N=2$ SUSY invariance, and dimensional power
counting}

As in the $N=1$ case, the $O(n)$ invariance of this measure
is obvious by inspection. As for $N=2$ \susy\ invariance under
\susyalgebratwo{},  this too is obvious for $\sum_{i=1,2}\xi_i Q_i$.
For $\sum_{i=1,2}\xibar_i\Qbar_i$ one can show that the arguments
of the four $\delta$-functions in \dmudeftwo\ transform into linear
combinations of one another, and furthermore that the associated 
transformation matrix has
superdeterminant unity, so that $N=2$ \susy\ is guaranteed. As in the
$N=1$ case this feature stems from the observation that these four
constraints assemble into a single $N=2$ multiplet as per Eq.~\Superadhm.

Next we verify the counting requirement  (iii) given in Sec.~2.2. 
Of the $2n^2+6n$ fermionic  modes in
\dmudeftwo, $2n(n-1)$ of them
are saturated by the second and third \hbox{$\delta$-functions}, 
leaving $8n$ unbroken
 adjoint fermion zero
modes as required. Note that the final $\delta$-function does not
lift any fermionic zero modes despite the presence of the fermion
bilinear $\Lambda_f$; this
is because the $\Atot$ integration  yields the purely
bosonic quantity $\det^{-1}\bigL$ as noted above. 
As in the $N=1$ theory, the bosonic dimensionality of \dmudeftwo\
scales, correctly, like 
\eqn\bosdimtwo{[a]^{\,(2n^2+6n)-(3n^2-3n)+(n^2-n)+(n^2-n)-(n^2-n)}\ \sim\ 
[a]^{8n}\ ,}
the five terms in the exponent coming, respectively, from the $w_i$
and $a'_{ij}$
integration variables, and from the four $\delta$-functions. (We do not
count factors of $\Atot$ here as they cancel between the differentials
and the $\delta$-functions.)

\subsec{Cluster decomposition}

Next we check cluster decomposition in the limit $|a'_{nn}|\rightarrow
\infty.$ The calculation proceeds just as in the $N=1$ case, except
that prior to eliminating the $\M'_{in}$ one eliminates the
$(\Atot)_{in}$ via
\eqn\Atotelim{\int\prod_{i=1}^{n-1}d(\Atot)_{in}\,
\delta\big(\,(\bigL\cdot\Atot-\Lambda
-\Lambda_f)_{i,n}\,\big)\,\times\,\cdots}
The argument of the $\delta$-function may be expanded as
\eqn\Atotarg{\big(\,\bigL\cdot\Atot-\Lambda
-\Lambda_f\big)_{i,n}\ =\ |a'_{nn}|^2(\Atot)_{in}+\cdots}
neglecting subleading terms. Thus these quantities scale like
\eqn\Atotscale{(\Atot)_{in}\ \sim\ |a'_{nn}|^{-2}}
and Eq.~\Atotelim\ yields a Jacobian factor
\eqn\Jatotdef{J_{\Atot}\ =\ 
\left({1\over|a'_{nn}|^2}\right)^{n-1}\ +\ \cdots,}
again neglecting subleading terms. The other new feature for the $N=2$ case
is the existence of the Higgsino collective coordinates $\N$. The
elimination of the $\N'_{in}$ proceeds identically to Eqs.~\subsetf-\Jacfdef\
for the $\M'_{in}$, and likewise gives
\eqn\JacfdefN{J_\N\ =\ |a'_{nn}|^{2(n-1)}\ +\ \cdots .}
The remaining integrations are just those of the $N=1$ model. Note
that the factors of $|a'_{nn}|$ again cancel among 
Eqs.~\Jacfdef, \measurehat, \Jadhmdef, \Jatotdef\ and \JacfdefN. 

It still
must be checked that as $|a'_{nn}|\rightarrow\infty$
 the arguments of the remaining $\delta$-functions in \dmudeftwo\
properly reduce to
those of the $(n-1)$-instanton case, i.e., are built from the truncated
matrices $\tilde a,$ $\tilde\M,$ $\tilde\N$ and $\tilde\A_{\rm tot}$ (see
Eq.~\atildedef). With the scaling relations \scalelike\ and \Atotscale\
this is easily verified, up to corrections of order $1/|a'_{nn}|^2.$
As in the $N=1$ case, the clustering condition \demand\ then gives, simply,
\eqn\Cnsolvetwo{C'_n\ =\ (C'_1)^n\ .}

\subsec{Agreement in the 2-instanton sector}

Still paralleling Sec.~2, we now check that in the 2-instanton sector,
our $O(n)$- and $N=2$ SUSY-invariant collective coordinate
measure \dmudeftwo\ and \Cnsolvetwo\ is equivalent to the
known first-principles $O(n)$-gauge-fixed measure \refs{\dkmone,\Osborn}
\eqn\dmuOsbtwo{\eqalign{\int\dmutwophys\ &=\ {1\over\Stwo}\int d^4w_1d^4w_2
d^4a'_{11}d^4a'_{22}d^2\mu_1d^2\mu_2d^2\M'_{11}d^2\M'_{22}
d^2\nu_1d^2\nu_2d^2\N'_{11}d^2\N'_{22}
\cr & \qquad\qquad\times\
\big(\,J_\bose/J_\fermi\,\big)^{1/2} \ .}}
The chief difference with the $N=1$ case is that now $\Jfermi$ is the
square of Eq.~\Jfermidef\ \dkmone, there being twice as many adjoint
fermions in the $N=2$ model. Consequently Eq.~\Jratio\ is modified to
\eqn\Jratiotwo{\big(\,J_\bose/J_\fermi\,\big)^{1/2}\ =\ C'_2\,
{\Big|\,|a_3|^2
\,-\,|a'_{12}|^2\,-\,\eighth{d\Sigma^\phi
\over d\phi}\big|_{\phi=0}\,\Big|\over H}\ ,}
where cluster still fixes
\eqn\newclust{C_2'\ =\ (C_1')^2}
as before. Now the insertions of unity into \dmuOsbtwo\
comprise, not just Eqs.~\unityb, \unityf, and \anotherone, but additionally
\eqn\unityftwo{1\ =\ {1\over4|a_3|^2}\int d^2\N'_{12}\,
\delta^2\big((\abar\N)_{1,2}-(\abar\N)_{2,1}\big)}
and
\eqn\unityAtot{1\ =\ \int d(\Atot)_{12}\,\delta\big(\,(\bigL\cdot\Atot-\Lambda
-\Lambda_f)_{1,2}\,\big)\,\det\bigL\ .}
Recall that $\bigL$ is a $\hf n(n-1)\,\times\,\hf n(n-1)$ dimensional
linear operator on the space of $n\times n$ antisymmetric matrices. For
$n=2$ this space is 1-dimensional, and $\bigL$ is simply the scalar $H$:
\eqn\Lbecomes{\det\bigL\ =\ \bigL\ = H\ .}
So the factor of $\det \bigL$ in Eq.~\unityAtot\
cancels the denominator in Eq.~\Jratiotwo. The rest of the argument
goes through precisely as in Sec.~2.5, and once again gives
exact equivalence to the postulated
$O(2)$- and SUSY-invariant form \dmudeftwo\ and
\Cnsolvetwo.

\newsec{Incorporation of matter}

\def\hyp{{\rm hyp}}

\def\Lambdahyp{{\Lambda_\hyp}}
The $N=1$ and $N=2$ multi-instanton measures detailed above for pure
$SU(2)$ gauge theory are easily extended to
incorporate fundamental matter multiplets. $N=1$ \susic\ fundamental
matter in the $n$-instanton sector is discussed in Appendix C of \dkmfour, and
in \oldyung.
For application to the
Section to follow, we will focus here on the $N=2$ case, and consider
adding
 $N_{F}$ matter hypermultiplets which transform in the fundamental
representation of $SU(2)$. Each $N=2$ hypermultiplet corresponds to a  
pair of
$N=1$ chiral multiplets, $Q_{i}$ and $\tilde{Q}_{i}$ where 
$i=1,2,\cdots, N_{F}$, which contain scalar quarks  $q_{i}$  
and
$\tilde{q}_{i}$ respectively and fermionic partners 
$\chi_{i}$ and $\tilde{\chi}_{i}$. 
 The fundamental fermion zero  modes associated with
$\chi_{i}$ and $\tilde{\chi}_{i}$
were constructed  in \refs{\CGTone,\Osborntwo}:
\eqn\fund{(\chi^{\alpha}_{i})^{\dot{\beta}}  \ = \
\bar{U}^{\dot{\beta}\alpha}_{\lambda}b_{\lambda k}f_{kl}{\cal K}_{li}
\ ,\quad
(\tilde{\chi}^{\alpha}_{i})^{\dot{\beta}}  \ = \
\bar{U}^{\dot{\beta}\alpha}_{\lambda}b_{\lambda
k}f_{kl}\tilde{\cal K}_{li}
}
with $\alpha$ a Weyl index and $\dot\beta$ an $SU(2)$ color index. 
Here each $\K_{ki}$ and  $\Kt_{ki}$
is a Grassmann number rather than a Grassmann spinor; there is no
$SU(2)$ index. The normalization matrix of these modes is given by  
\Osborntwo
\eqn\orth{\int\, d^{4}x (\chi^{\alpha}_{i})_{\dot{\beta}}
(\tilde{\chi}_{\alpha j})^{\dot{\beta}}\ = \
\pi^{2}{\cal K}_{li}\tilde{\cal K}_{lj} 
}
so that the hypermultiplet part of the $n$-instanton measure  reads
\dkmfour
\eqn\muhypdef{\int d\muhyp^{(n)}\ =\
{1\over\pi^{2nN_F}}\int\prod_{i=1}^{N_F}d\K_{1i}\cdots d\K_{ni}\,
d\Kt_{1i}\cdots d\Kt_{ni}\ .}
The total measure is then simply 
\eqn\totmeasure{\dmunphys\times d\muhyp^{(n)}\ .}

One can also consider the case of coupling to a single (massive) $N=2$
matter hypermultiplet in the adjoint representation of the gauge group,
but this is best understood by starting from the $N=4$ theory and
will be discussed in a separate publication.

\newsec{Explicit expressions for the Seiberg-Witten prepotentials,
and for the \hbox{$4$-derivative/$8$-fermion} term}

In earlier work \dkmfour\ we presented a general formula for the
$\F_n$ (i.e., the $n$-instanton contribution to the Seiberg-Witten
prepotentials \refs{\SWone,\SWtwo}) as integrations of the exponentiated
multi-instanton action over the space
of  $N=2$ \susic\ collective coordinates. However,
at the time, the collective coordinate integration measure was not known. For
the purposes of self-containedness, we repeat those expressions here,
inserting the explicit expression for the $N=2$ measure, Eqs.~\dmudeftwo\ and
\Cnsolvetwo.

We start with the case of pure $N=2$ \susic\ $SU(2)$ gauge theory
\refs{\SWone,\dkmone}. In this model the $n$-instanton action 
reads\foot{This expression contrasts with the $N=1$ \susic\ action,
which in ADHM variables looks like the noninteracting sum of $n$
single instantons; see Appendix C of \dkmfour, and Ref.~\oldyung.}
 \dkmone
\def\Lambdabar{\bar\Lambda}

\eqn\sinstfinal{S^0_{\rm inst}\  =\
{8n\pi^2\over g^2}\ +\ 16\pi^2|\vhiggsa|^2\sum_{k=1}^n|w_k|^2
\ -\ 8\pi^2\,\Tr_n\,\Atot\Lambdabar
\ +\   
4\sqrtwo\pi^2\,\mu_k^\alpha\,\barvhiggsa{}_\alpha{}^\beta\,\nu_{k\beta}\ 
 ,}
where the notation is that of Eqs.~\bcanonical, \Mdef, \Ndef,
and \vevbcagain-\Lambdabardef\ above.  As shown in \dkmfour, this may
be rewritten as a manifestly $N=2$ \susic\ ``$F$-term''$\,$:
\eqn\Scompact{S^0_{\rm inst}\ =\ {8n\pi^2\over g^2}\ 
-\ \pi^2\,\Tr\,\abar(\thetabar_i)\big(\Pinfty+1\big)a(\thetabar_i)\
\Big|_{\thetabar_1^2\thetabar_2^2}\ .}
Here the capitalized `Tr' indicates a trace over both ADHM and $SU(2)$
indices, $\Pinfty$ denotes  the  
$(n+1)\times(n+1)$ matrix
\eqn\Pinftydef{\Pinfty\ =\  =\ 1-b\bbar
\ =\ \delta_{\lambda0}\delta_{\kappa0}\ ,}
and $a(\thetabar_i)$ is the space-time-constant ``superfield''
given in Eq.~\apromote.

As for the measure, it is useful to factor the physical $N=2$ measure,
Eq.~\dmudeftwo\ and \Cnsolvetwo, as follows:
\def\dmunphystilde{d\tilde\mu^{(n)}_{\rm phys}}
\eqn\mufactor{\int\dmunphys\ =\ \int d^4x_0\,d^2\xi_1\,d^2\xi_2\,
\int\dmunphystilde}
where $(x_0,\xi_1,\xi_2)$ gives the global position of the 
multi-instanton in $N=2$ superspace. Explicitly, $x_0,$ $\xi_1$ and
$\xi_2$ are the linear combinations proportional to the `trace' components
of the $n\times n$ matrices $a',$ $\M'$ and $\N'$, respectively \dkmone:
\eqn\traceparts{x_0\ =\ {1\over n}\Tr_n\,a'\ ,\quad
\xi_1\ =\ {1\over4n}\Tr_n\M'\ ,\quad
\xi_2\ =\ {1\over4n}\Tr_n\N'\ .}
Note that these $N=2$ superspace modes do not enter into the
$\delta$-function constraints in \dmudeftwo\
and so do indeed factor out in this simple way. Furthermore, the four
exact \susic\ modes $\xi_{1\alpha}$ and 
$\xi_{2\alpha}$ are the only fermionic modes that are not lifted by
(i.e., do not appear in) the action \sinstfinal.

In terms of these quantities, the exact all-instanton-orders expression for 
the Seiberg-Witten prepotential reads:
\eqn\FSWdef{\F(\Nequalstwo)\ \equiv
\ {\cal F}_{\rm pert}(\Nequalstwo)\,+\,{\cal F}_{\rm inst}(\Nequalstwo)
\ =\
{i\over2\pi}
\Nequalstwo^2\,\log{2\Nequalstwo^2\over e^3\Lambda^2}\ -\ {i\over\pi}
\sum_{n=1}^\infty
{\cal F}_n\,\left({\Lambda\over\Nequalstwo}\right)^{4n}\Nequalstwo^2\ .}
Here $\Lambda$ is the  dynamically generated scale in the Pauli-Villars
scheme, and the $\F_n$ may be expressed as
the following explicit finite-dimensional collective coordinate
integrals \dkmfour:
\eqn\Fnfinal{\F(\vhiggs){{}\atop\Big|}_{n\hbox{-}{\rm inst}}\ 
\equiv\ -{i\F_n\over\pi}
\left({\Lambda\over\Nequalstwo}\right)^{4n}\Nequalstwo^2
\ =\ 8\pi i\int\dmunphystilde\,\exp(-S^0_{\rm inst})\ .}

Next we consider the class of models in which $N=2$ \susic\ $SU(2)$
gauge theory is coupled to $N_F$ flavors of massless quark hypermultiplets
\refs{\SWtwo,\dkmfour}. In these cases the $n$-instanton action
reads \dkmfour
\def\SNF{S^{N_F}_{\rm inst}}
\def\Szero{S^{0}_{\rm inst}}
\eqn\SNFfinal{\SNF\ =\ \Szero\ -\ 8\pi^2\Tr_n\,\Lambdahyp\,\Atot\ ,}
where $\Lambdahyp$ is the $n\times n$ antisymmetric matrix
\eqn\Lambdahypdef{(\Lambda_\hyp)_{k,l}\ =\ {i\sqrtwo\over16}\,
\sum_{i=1}^{N_F}\big(\K_{ki}\Kt_{li}+\Kt_{ki}\K_{li}\big)\ ,}
and $\K$ and $\Kt$ are the fundamental fermion collective coordinates
defined in Eq.~\fund. 
In these models the all-instanton-orders contributions to the prepotential
are still correctly given by Eq.~\Fnfinal, with the substitutions
\eqn\substa{\Szero\ \rightarrow\ \SNF\ ,\qquad 
\dmunphystilde\ \rightarrow\ \dmunphystilde\,\times\,d\muhyp^{(n)}\ }
with $d\muhyp^{(n)}$ as in Eq.~\muhypdef. Actually for $N_F>0$
 all odd-instanton contributions vanish in the massless case
due to a discrete $\bigZ_2$ symmetry \refs{\SWtwo,\dkmfour}.
The incorporation of hypermultiplet masses is straightforward but rather too
lengthy to recapitulate here; see Ref.~\dkmfour\ for a discussion of
the $\F_n$ in this case, and (say) Sec.~2.1 of 
\dkmfive\ for the full expression
for the perturbative part of $\F$ (which is often incompletely rendered
in the literature). 

Given these expressions for the prepotential, one also knows the
all-instanton-orders expansion of the quantum modulus
$u=\langle\Tr\Phi^2\rangle$, since on general grounds 
\refs{\Matone-\BMSTY}
\eqn\matrelis{u(\vhiggs){{}\atop\Big|}_{n\hbox{-}{\rm inst}}\ =\ 2i\pi n\cdot
\F(\vhiggs){{}\atop\Big|}_{n\hbox{-}{\rm inst}}\ .}

The above collective coordinate integral expressions for $\F$ and $u$
constitute
a closed solution, in quadratures, of the low-energy dynamics of the
Seiberg-Witten models. It is interesting that this solution
is obtained purely from the semiclassical regime, without appeal
to electric-magnetic duality. For $N_F\le3$ this may be regarded as
academic, in light of the exact results of \refs{\SWone,\SWtwo}. 
However, for the conformally invariant case $N_F=4$, the
multi-instanton solution contains information
not present in \SWtwo, namely the all-instanton-orders relation between
the microscopic ($SU(2)$) and effective ($U(1)$)
complexified coupling constants,
$\tau_{\rm micro}$ and $\tau_{\rm eff}$.  
For $N_F=4,$ Eq.~\FSWdef\ should simply be replaced by 
\eqn\newFSW{\F(\vhiggs)\ =\ \quarter\, \tau_{\rm micro}\,\vhiggs^2
\ -\ {i\over\pi}\sum_{n=0,2,4,\cdots}\F_n\,q^n\,\vhiggs^2}
where $q=\exp(i\pi\tau_{\rm micro}),$ and the $\F_n$ have the same
collective coordinate integral representation  as for $N_F<4.$ 
By definition, the effective coupling $\tau_{\rm eff}$ is obtained
from the prepotential via $\tau_{\rm eff}=2\F''(\vhiggs)$.
The constants $\F_0$ and $\F_2$ were explicitly evaluated in
Refs.~\dkmfive\ and \dkmfour, respectively, and were found to be nonzero.
We do not see how such a relation between
$\tau_{\rm micro}$ and $\tau_{\rm eff}$  
could be obtained using the
methods of \SWtwo, since the modular group ostensibly
acts only on $\tau_{\rm eff}$ and not on $\tau_{\rm micro}$. Nevertheless,
it may be that the series connecting them sums to a modular function, so
that the modular group also acts on 
 $\tau_{\rm micro}$.

Of course, the prepotential only controls the leading
2-derivative/4-fermion terms in the gradient expansion along the
Coulomb branch of $N=2$ \susic\ QCD. In general one has
\def\Ltwoderiv{{\cal L}_{2\hbox{-}\rm deriv}}
\def\Lfourderiv{{\cal L}_{4\hbox{-}\rm deriv}}
\def\Psibar{\bar\Psi}
\eqn\Lgraddef{{\cal L}_{\rm eff}\ =\ \Ltwoderiv\ +\ \Lfourderiv\ + \cdots}
where
\eqn\Ltwoderivdef{\Ltwoderiv\ =\ {1\over4\pi}\,
{\rm Im}\int d^4\theta\,\F(\Psi)\ ,}
and
\eqn\Lfourderivdef{\Lfourderiv\ =\ \int d^4\theta d^4\bar\theta\,
\H(\Psi,\Psibar)\ .}
Here $\Psi$ is the $N=2$ chiral superfield, and
$\H$ is a real function of its arguments \refs{\Henningson,\dWGR}.
The pure $n$-instanton contribution to $\H$ for pure $SU(2)$ gauge
theory, valid to leading semiclassical order, is then given by:
\eqn\Hndef{{\partial^4\over\partial\vbarhiggs^4}\,\H(\vhiggs,\vbarhiggs)\,
{\Big|}_{n\hbox{-}\rm inst}\ =\ 64 \pi^8
\int d\tilde\mu_{\rm phys}^{(n)}\,e^{-S_{\rm inst}^{0}}\,
\sum_{k,k',l,l'=1}^n\,
(\nu^{}_k\tau^3w^{}_k\wbar^{}_{k'}\tau^3\nu^{}_{k'}) 
(\mu^{}_l\tau^3w^{}_l\wbar^{}_{l'}\tau^3\mu^{}_{l'})}
where the VEVs $\vhiggs$ and $\vbarhiggs$ are to be treated as independent.
This expression was written down in Ref.~\dkmnine; the new information
here is the explicit definition of the collective coordinate measure
$d\tilde\mu_{\rm phys}^{(n)},$ defined by Eqs.~\mufactor, \dmudeftwo\ and
\Cnsolvetwo. For the pure \hbox{$n$-antiinstanton} contribution, exchange
$\vhiggs$ and $\vbarhiggs,$ while for $N_F>0$ make the changes \substa.
As emphasized in \dkmnine, in contrast to the nonrenormalized
holomorphic prepotential, and excepting the special case $N_F=4$
\DineSeib,
there will in general be perturbative corrections to Eq.~\Hndef\ as well
as mixed $n$-instanton, $m$-antiinstanton contributions to $\H$
not governed by Eq.~\Hndef.

We close with a practical 
comment on the doability of these integrations for the
prepotentials.  Notice that when $\Atot$ is eliminated at the outset 
in favor
of the other moduli via the invertible linear equation \thirtysomething,
then the action \sinstfinal\ becomes a complicated algebraic expression,
and the resulting integrations are quite involved; only numerical methods
appear promising for $n>2$. On the other hand, if
 $\Atot$ is treated
as an independent integration variable (as we have been doing throughout), then
both the $n$-instanton action \sinstfinal\ and \SNFfinal, 
and the arguments of the constraint
$\delta$-functions in \dmudeftwo, are quadratic forms in the collective
coordinates $\{a,\M,\N,\K,\Kt\}$. (In fact the action only involves
the top-row elements of $a,$ $\M$ and $\N.$) It follows that the
above-given
expressions for the $\F_n$ are
amenable to standard methods of analysis
for (supersymmetric)
Gaussian integrals. For instance, if one were to exponentiate
the constraints in the usual way, by means of a supermultiplet of
Lagrange multipliers, then the Gaussian variables
$\{a,\M,\N,\K,\Kt\}$ can be integrated out entirely, and replaced by the
appropriate superdeterminant. Only the $\Atot$ integration, and the
integration over the Lagrange multipliers, 
remains. It would be interesting
if such an expression would naturally yield a recursion formula in $n$
for the $\F_n$, such as Matone's \Matone\ for instance.

\listrefs
\bye